\documentclass[useAMS,usenatbib]{mn2e}

\usepackage{graphicx}
\usepackage{times}%
\usepackage{natbib}
\usepackage{graphics}
\usepackage{epsfig}
\usepackage{array}
\usepackage{stfloats}
\usepackage{fixltx2e}
\usepackage{amsmath}

\newcommand{\ie}{{i.e.}}
\newcommand{\eg}{{e.g.}}
\newcommand{\gsim}{\,\lower2truept\hbox{${>\atop\hbox{\raise4truept\hbox{$\sim$}}}$}\,}

\def\eg{{\rm e.g.$\,$}}
\def\ie{{\rm i.e.$\,$}}

\newcommand{\be}{\begin{equation}}
\newcommand{\ee}{\end{equation}}
\newcommand{\bea}{\begin{eqnarray}}
\newcommand{\eea}{\end{eqnarray}}

\renewcommand{\vec}[1]{ {\bmath #1} } 

\def\ltsima{$\; \buildrel < \over \sim \;$}
\def\simlt{\lower.5ex\hbox{\ltsima}}
\def\gtsima{$\; \buildrel > \over \sim \;$}
\def\simgt{\lower.5ex\hbox{\gtsima}}

\title[The CoDECS project]{The CoDECS project: a publicly available suite of cosmological N-body simulations for interacting dark energy models}

\author[M. Baldi]{Marco Baldi
\\Excellence Cluster Universe, Boltzmannstr.~2, D-85748 Garching, Germany
\\University Observatory, Ludwig-Maximillians University Munich, Scheinerstr. 1, D-81679 Munich, Germany
\\Email: marco.baldi@universe-cluster.de
\\{\em Simulations publicly available at: www.marcobaldi.it/research/CoDECS}}

\begin{document}


\pagerange{\pageref{firstpage}--\pageref{lastpage}} \pubyear{2011}

\maketitle

\label{firstpage}

\begin{abstract}	
\ \\
We present the largest set of N-body and hydrodynamical simulations to date for cosmological models featuring
a direct interaction between the Dark Energy (DE) scalar field, responsible of the observed cosmic acceleration, and
the Cold Dark Matter (CDM) fluid. With respect to previous works, our simulations considerably extend the statistical significance 
of the simulated volume and cover a wider range of different realizations of the interacting DE scenario, 
including the recently proposed {\em bouncing} coupled DE model. 
Furthermore, all the simulations are normalized in order to be consistent with 
the present bounds on the amplitude of density perturbations at last scattering, thereby providing the first
realistic determination of the effects of a DE coupling for cosmological growth histories fully compatible
with the latest Cosmic Microwave Background data. 
As a first basic analysis, we have studied the impact of the coupling on the nonlinear
matter power spectrum and on the bias between the CDM and baryon distributions, as a function of redshift and scale.
For the former, we have addressed the issue of the degeneracy between the effects of the coupling and other standard cosmological
parameters, as \eg $\sigma _{8}$, showing how the redshift evolution of the linear amplitude or the scale dependence of the nonlinear power spectrum
might provide a way to break the degeneracy. For the latter, instead, we have computed the redshift and scale dependence of the bias
in all our different models showing how a growing coupling or a {\em bouncing} coupled DE scenario provide much stronger effects with respect to
constant coupling models. Furthermore, we discuss the main features imprinted by the DE interactions on the
Halo and Subhalo Mass Functions.
We refer to this vast numerical initiative as the COupled Dark Energy Cosmological Simulations project, or {\small CoDECS},
and we hereby release all the {\small CoDECS} outputs for public use through a dedicated web database, providing information
on how to access and interpret the data.

\end{abstract}



\section{Introduction}
\label{i}

The observational evidence in favor of a relatively recent acceleration of the expansion of the Universe has been progressively reinforced by a large number of
independent probes \citep[see \eg][]{SNLS,Kowalski_etal_2008,Percival_etal_2001,Cole_etal_2005,Percival_etal_2010,Reid_etal_2010} 
over more than a decade since the first discovery of the dimming of distant Type Ia Supernovae \citep[][]{Riess_etal_1998,Perlmutter_etal_1999}. 
The fundamental origin of such acceleration -- generically referred to as the ``Dark Energy" -- represents one of the central open problems in modern cosmology, and is the main target of a wide range
of ambitious observational initiatives planned for the upcoming future such as PanStarrs \citep[][]{PanStarrs}, HETDEX \citep[][]{HETDEX}, DES \citep[][]{DES},
LSST \citep[][]{LSST} and {\em Euclid}\footnote{http://www.euclid-ec.org} \citep[][]{EUCLID-r}. 

The standard $\Lambda $CDM cosmological model accounts for the observed Dark Energy by means of a cosmological constant $\Lambda $,
\ie a completely homogeneous form of energy whose density does not dilute with the expansion of the universe. Such a simple prescription is
sufficient to fit the vast majority of presently available data, but requires an unnatural level of fine-tuning of the cosmological constant value, 
giving rise to the so-called ``fine-tuning" and ``coincidence" problems of the standard $\Lambda $CDM cosmology \citep[see eg. ][]{Weinberg_1988}.
For this reason, alternative Dark Energy scenarios have been proposed, including models based on dynamical scalar fields \citep[][]{Wetterich_1988,Ratra_Peebles_1988,kessence}
or on modifications of General Relativity at cosmological scales \citep[see \eg ][]{Hu_Sawicki_2007}.

Furthermore, a wide range of astrophysical data at intermediate and small scales seem to show possible deviations from the expectations of the $\Lambda $CDM
model, thereby providing additional motivation for the exploration of alternative scenarios. Such deviations range from the lack of luminous satellites
in CDM halos \citep[the so-called ``satellite problem", ][]{Navarro_Frenk_White_1995}, to the observed low baryon fraction in galaxy clusters \citep[see \eg ][]{Ettori_2003,Allen_etal_2004,Vikhlinin_etal_2006,
LaRoque_etal_2006,McCarthy_etal_2007}, to the shallow observed CDM density profiles in dwarf galaxies \citep[the so-called ``cusp-core" problem, ][]{Moore_1994,Flores_Primack_1994,Simon_etal_2003}, in spiral galaxies \citep[][]{Navarro_Steinmetz_2000,Salucci_Burkert_2000,Salucci_2000,Binney_Evans_2001}, 
and in galaxy clusters \citep[][]{Sand_etal_2002,Sand_etal_2004,Newman_etal_2009}, to the anomalous detection of massive clusters at high redshift \citep[][]{Mullis_etal_2005,Bremer_etal_2006,Jee_etal_2009,Rosati_etal_2009,Brodwin_etal_2010,Jee_etal_2011,Foley_etal_2011}, and to the high velocities
detected in the large-scale bulk motion of galaxies \citep[\eg][]{Watkins_etal_2009} or in systems of colliding galaxy clusters as the ``Bullet Cluster" \citep[][]{Lee_Komatsu_2010,Lee_2010}.

Most of these anomalous observations involve highly nonlinear environments where complex astrophysical processes play a major role in shaping the evolution of structures and 
of their direct observable properties. It is therefore reasonable to investigate a possible astrophysical origin of such deviations from the expectations of an idealized
$\Lambda $CDM cosmology by refining our modeling of all the relevant astrophysical processes in place at the scales of interest for these different observations, as 
\eg gas cooling, star formation, and feedback mechanisms from supernovae explosions in galaxies or from Active Galactic Nuclei in galaxy clusters.
Nevertheless, such deviations could also represent the signature of the underlying cosmological model, thereby providing a unique handle to investigate alternative scenarios to the
standard $\Lambda $CDM cosmology and to constrain the properties of the dark sector of the universe.

In both cases, however, due to the high nonlinearity of the systems under investigation, it is necessary to rely on complex and computationally demanding numerical simulations 
where the specific astrophysical or cosmological models to be studied have to be properly and efficiently implemented in order to provide reliable predictions to be compared with
real astrophysical data. In this respect, if on one side a great effort has been made in order to refine the theoretical modeling and the numerical implementation of complex astrophysical
processes in cosmological N-body codes \citep[see \eg][]{Springel_Hernquist_2003b,Maio_etal_2007,Sijacki_etal_2007,Springel_2011}, not as much deal of work has been put into the modification of standard N-body algorithms in order to account for
the nonlinear evolution of alternative cosmological scenarios. In the last years, some first steps have been moved in this direction by a few numerical studies of non-standard
cosmological scenarios as \eg Modified Gravity models \citep[in the form of $f(R)$ theories, ][]{Oyaizu_2008,Oyaizu_etal_2008,Schmidt_etal_2009,Zhao_Li_Koyama_2011}, Warm Dark Matter models
\citep[see \eg][]{Zavala_etal_2009,Lovell_etal_2012,Viel_etal_2011}, Extended Quintessence models \citep[][]{DeBoni_etal_2010,Li_Mota_Barrow_2011}, or Coupled Dark Energy models \citep[][]{Maccio_etal_2004,Baldi_etal_2010,Li_Barrow_2011}.

In particular, the latter have been shown to determine significant deviations from the expectations of $\Lambda $CDM in \eg the predicted baryon fraction and concentration of galaxy and cluster halos
\citep[][]{Baldi_etal_2010,Baldi_2011a}, the predicted transmitted flux of Lyman-$\alpha $ systems \citep[][]{Baldi_Viel_2010}, the number counts of massive galaxy clusters
as a function of redshift \citep[][]{Mainini_Bonometto_2006,Baldi_Pettorino_2011,Baldi_2011c,Tarrant_etal_2012}, and the distribution of galaxies within a galaxy cluster main halo \citep[][]{Baldi_Lee_Maccio_2011}.

Coupled Dark Energy models have been first proposed by \citet{Wetterich_1995} as a possible way to alleviate the cosmological ``fine-tuning" problem, and subsequently studied by 
several authors \citep[see \eg][]{Amendola_2000,Amendola_2004,Farrar2004,Mainini_Bonometto_2006,Farrar2007,Pettorino_Baccigalupi_2008,Amendola_Baldi_Wetterich_2008,Koyama_etal_2009,CalderaCabral_2009,
Honorez_etal_2010,Baldi_2011a,Baldi_2011c}. These are based on the dynamical evolution of a classical scalar field $\phi $ that plays the role of the Dark Energy,
and feature a direct interaction between this Dark Energy scalar field and the Dark Matter fluid, that exchange energy-momentum during the cosmic evolution.
Such interaction gives rise to a ``fifth-force" between Dark Matter particles -- mediated by the Dark Energy scalar field -- that significantly modifies the gravitational instability processes through which
cosmic structures develop, both in the linear and in the nonlinear regimes.
A detailed investigation of the nonlinear evolution of this particular class of alternative cosmological models is therefore necessary to provide a quantitative determination of the deviations
expected from the standard $\Lambda $CDM predictions in several observational contexts, and to assess the viability of the Coupled Dark Energy scenario as a possible
alternative to the standard model.

In this work we will present the largest and most accurate N-body simulations to date for Coupled Dark Energy cosmologies in terms of simulated volume, numerical resolution, and
range of models covered in the numerical sample. These include both collisionless runs at large scales and adiabatic hydrodynamical simulations at small scales for five different
Coupled Dark Energy scenarios, besides the standard fiducial $\Lambda $CDM cosmology. The various Coupled Dark Energy models include constant coupling models \citep[][]{Amendola_2000},
variable coupling models \cite[][]{Baldi_2011a}, and the recently proposed Bouncing Coupled dark Energy scenario \citep[][]{Baldi_2011c}. 
All the models share the same set of cosmological parameters at the present time, and
the same amplitude of density perturbations at the redshift of the last scattering surface ($z_{\rm CMB}\approx 1100$), both consistent with the latest results from the WMAP satellite \citep[][]{wmap7}.
The present set of cosmological simulations is therefore fully compatible with CMB constraints and provides a detailed picture of the evolution of the universe from the last scattering surface to the present epoch
within five different Coupled Dark Energy scenarios.

We will therefore illustrate the details of this vast numerical study 
that goes under the name of  the ``{\small CoDECS} project", where the acronym {\small CoDECS} stands for (COupled Dark Energy Cosmological Simulations).
All the simulations outputs -- that include, besides the raw snapshot files, also halo and subhalo catalogs and the nonlinear matter power spectra -- are made publicly available through
a dedicated web database together with the background and linear perturbations evolution computed for each different model.\\

The paper is organized as follows. In Section~\ref{models} we review the basic equations and main features of Couped Dark Energy models both concerning their background and liner perturbations evolution.
In Section~\ref{sec:sims} we introduce our set of N-body simulations, describing their numerical features and the procedures adopted to generate initial conditions, and comparing the
visual outcomes of a few sample simulations for different Coupled Dark Energy scenarios. In Section~\ref{sec:results} we present some basic results derived from the analysis of our large-scale
collisionless simulations, focusing on the time evolution of the nonlinear matter power spectrum and on the offset between the baryonic and the Cold Dark Matter overdensities. 
We also provide a few examples of the possible applications of the {\small CoDECS} halo catalogs by computing the Halo Mass Function and the Subhalo Mass Function for the different cosmologies under investigation, and highlighting the main characteristic features imprinted by the Dark Eenergy interaction on these observable quantities.
Finally, in Section~\ref{sec:concl} we draw our conlusions. 

Furthermore, in the Appendix we provide the details of the public database where the simulations outputs are stored and accessible for public use, 
briefly describing the contents of the database and how different information are organized in different files. High-resolution versions of the images displayed in this paper can also be downloaded from the
same website. 

\section{Interacting Dark Energy Models}
\label{models}

The {\small CoDECS} project is aimed at providing publicly available data from large N-body simulations
for a significant number of interacting Dark Energy (DE) cosmological models. 
Interacting DE scenarios have been extensively studied in the last decade as a possible alternative to the
standard $\Lambda $CDM cosmology as they feature a dynamical evolution of the DE density that provides
a way to partially address the fine-tuning problems that affect the cosmological constant \citep[see \eg][]{Weinberg_1988} and to simultaneously
account for an accelerated expansion of the universe at late times.
Starting from the initial proposal of \citet{Wetterich_1995}, this class of cosmological models has been investigated by several authors
\citep{Amendola_2000, Amendola_2004,Pettorino_Baccigalupi_2008,Amendola_Baldi_Wetterich_2008,CalderaCabral_2009,Koyama_etal_2009} to which we refer for a detailed derivation of the main equations and for an
extensive discussion of the main features of coupled DE (cDE, hereafter) cosmologies. 

In the present Section we will only briefly review the basic equations that characterize cDE scenarios and we will present
the specific models that are included in our suite of N-body simulations, providing a direct comparison of their
background and linear perturbations evolution.

\subsection{Background evolution}

We consider a set of flat cosmological models that include CDM, baryons, radiation and a classical
DE scalar field $\phi $ (subscripts $c$, $b$, $r$, and $\phi $, respectively). 
For any given set of cosmological parameters $\Omega _{r}$, $\Omega _{b}$, $\Omega _{\phi }$,
flatness is ensured by imposing $\Omega _{c}=1-\Omega _{r}-\Omega _{b}-\Omega _{\phi }$. In particular, for all the models
included in the {\small CoDECS} suite we consider cosmological parameters in agreement with the ``WMAP7 only Maximum Likelihood"
results of \citet{wmap7}, that are listed in Table~\ref{tab:parameters}. 
\begin{table}
\begin{center}
\begin{tabular}{cc}
\hline
Parameter & Value\\
\hline
$H_{0}$ & 70.3 km s$^{-1}$ Mpc$^{-1}$\\
$\Omega _{\rm CDM} $ & 0.226 \\
$\Omega _{\rm DE} $ & 0.729 \\
${\cal A}_{s}$ & $2.42 \times 10^{-9}$\\
$ \Omega _{b} $ & 0.0451 \\
$n_{s}$ & 0.966\\
\hline
\end{tabular}
\end{center}
\caption{The set of cosmological parameters at $z=0$ assumed for all the models included in the {\small CoDECS} project, consistent with the latest 
results of the WMAP collaboration for CMB data alone \citep[][]{wmap7}.}
\label{tab:parameters}
\end{table}
\begin{table*}
\begin{tabular}{llcccccccc}
Model & Potential  &  
$\alpha $ &
$\beta _{0}$ &
$\beta _{1}$ &
\begin{minipage}{45pt}
Scalar field \\ normalization
\end{minipage} &
\begin{minipage}{45pt}
Potential \\ normalization
\end{minipage} &
$w_{\phi }(z=0)$ &
${\cal A}_{s}(z_{\rm CMB})$ &
$\sigma _{8}(z=0)$\\
\\
\hline
$\Lambda $CDM & $V(\phi ) = A$ & -- & -- & -- & -- & $A = 0.0219$ & $-1.0$ & $2.42 \times 10^{-9}$ & $0.809$ \\
EXP001 & $V(\phi ) = Ae^{-\alpha \phi }$  & 0.08 & 0.05 & 0 & $\phi (z=0) = 0$ & $A=0.0218$ & $-0.997$ & $2.42 \times 10^{-9}$ & $0.825$ \\
EXP002 & $V(\phi ) = Ae^{-\alpha \phi }$  & 0.08 & 0.1 & 0 &$\phi (z=0) = 0$ & $A=0.0218$ & $-0.995$ & $2.42 \times 10^{-9}$ & $0.875$ \\
EXP003 & $V(\phi ) = Ae^{-\alpha \phi }$  & 0.08 & 0.15 & 0 & $\phi (z=0) = 0$ & $A=0.0218$ & $-0.992$ & $2.42 \times 10^{-9}$ & $0.967$\\
EXP008e3 & $V(\phi ) = Ae^{-\alpha \phi }$  & 0.08 & 0.4 & 3 & $\phi (z=0) = 0$ & $A=0.0217$ & $-0.982$ & $2.42 \times 10^{-9}$ & $0.895$ \\
SUGRA003 & $V(\phi ) = A\phi ^{-\alpha }e^{\phi ^{2}/2}$  & 2.15 & -0.15 & 0 & $\phi (z\rightarrow \infty ) = \sqrt{\alpha }$ & $A=0.0202$ & $-0.901$ & $2.42 \times 10^{-9}$ & $0.806$ \\
\hline
\end{tabular}
\caption{The list of cosmological models considered in the {\small CoDECS} project and their specific parameters. The scalar field is normalized to be zero
at the present time for all the models except the bouncing cDE scenario, for which the normalization is set in the very early universe by placing the field 
at rest in its potential minimum \citep[see][for further details]{Baldi_2011c}. All the models have the same amplitude of scalar perturbations at $z_{\rm CMB}\approx 1100$,
as shown by the common value of the amplitude ${\cal A}_{s}$, but have very different values of $\sigma _{8}$ at $z=0$, again with the sole exception of the bouncing cDE
model SUGRA003.}
\label{tab:models}
\end{table*}

The background dynamic equations for the different cosmic components are given by:
\begin{eqnarray}
\label{klein_gordon}
\ddot{\phi } + 3H\dot{\phi } +\frac{dV}{d\phi } &=& \sqrt{\frac{2}{3}}\beta _{c}(\phi ) \frac{\rho _{c}}{M_{{\rm Pl}}} \,, \\
\label{continuity_cdm}
\dot{\rho }_{c} + 3H\rho _{c} &=& -\sqrt{\frac{2}{3}}\beta _{c}(\phi )\frac{\rho _{c}\dot{\phi }}{M_{{\rm Pl}}} \,, \\
\label{continuity_baryons}
\dot{\rho }_{b} + 3H\rho _{b} &=& 0 \,, \\
\label{continuity_radiation}
\dot{\rho }_{r} + 4H\rho _{r} &=& 0\,, \\
\label{friedmann}
3H^{2} &=& \frac{1}{M_{{\rm Pl}}^{2}}\left( \rho _{r} + \rho _{c} + \rho _{b} + \rho _{\phi} \right)\,,
\end{eqnarray}
where the source terms at the right hand side of Eqs.~(\ref{klein_gordon},\ref{continuity_cdm}) encode the 
interaction between the DE scalar field $\phi $ and CDM particles. 
In Eqs.~(\ref{klein_gordon},\ref{continuity_cdm}), and throughout the paper, 
the scalar field $\phi $ is expressed in units of the reduced Planck mass $M_{{\rm Pl}}\equiv 1/\sqrt{8\pi G}$
and an overdot denotes a derivative with respect to the cosmic time $t$.
The coupling function $\beta _{c}(\phi )$
sets the strength of the interaction while the sign of the quantity $\dot{\phi }\beta _{c}(\phi )$ 
determines the direction of the energy-momentum flow between
the two components. With the convention assumed in Eqs.~(\ref{klein_gordon},\ref{continuity_cdm}) a positive 
combination $\dot{\phi }\beta _{c}(\phi ) > 0$ corresponds to a transfer of energy-momentum from CDM to DE, while the opposite
trend occurs for negative values of $\dot{\phi }\beta _{c}(\phi )$. 
Such energy transfer implies a time variation of the CDM particle mass, according to the equation:
\begin{equation}
\label{mass_variation}
m_{c}(z)=m_{c,0}\cdot e^{-\int \beta _{c}(\phi )d\phi }
\end{equation}
which can be obtained by integrating Eq.~(\ref{continuity_cdm}), that implies $\dot{m}_{c} < 0$ for $\dot{\phi }\beta _{c}(\phi ) > 0$
and $\dot{m}_{c} > 0$ for $\dot{\phi }\beta _{c}(\phi ) < 0$.
As shown by Eqs.~(\ref{continuity_baryons},\ref{continuity_radiation}),
baryons and radiation are always uncoupled from the DE scalar field, such that $\rho _{b}\propto a^{-3}$ and $\rho _{r} \propto a^{-4}$
for all the cosmologies considered in the present work, 
and Eq.~(\ref{friedmann}) closes the system by imposing flatness.
\ \\

In the range of models included in the {\small CoDECS} project we will consider two possible choices for the scalar field self-interaction potential $V(\phi )$, namely
an exponential potential \citep{Wetterich_1988}:
\begin{equation}
\label{exponential}
V(\phi ) = Ae^{-\alpha \phi }
\end{equation}
and a SUGRA potential \citep{Brax_Martin_1999}:
\begin{equation}
\label{SUGRA}
V(\phi ) = A\phi ^{-\alpha }e^{\phi ^{2}/2} \,.
\end{equation}
The former is known to provide stable scaling solutions for the scalar field that are almost independent from the field's initial conditions 
even for an uncoupled system \citep[\ie for $\beta _{c}(\phi ) = 0$, see \eg][]{Ferreira_Joyce_1998}. For the case of coupled DE ($\beta _{c}(\phi )\neq 0$),
instead, it features a transient early DE scaling solution (known as the $\phi $-Matter Dominated Epoch, or $\phi$MDE) followed by a late-time
accelerated attractor \citep[see \eg][]{Amendola_2000, Amendola_2004,Baldi_2011a}. 
The latter potential form, instead, naturally arises in the context of supersymmetric theories of gravity and provides a viable cosmological evolution
with a late-time accelerated attractor even in the absence of any coupling, while for the case of a cDE scenario it has been recently shown 
\citep{Tarrant_etal_2012,Baldi_2011c} to have a significant impact on the evolution of density perturbations.
In particular, \citet{Baldi_2011c} has extensively investigated the same SUGRA cDE model presented in this work, 
showing that it determines a ``bounce" of the DE equation of state $w_{\phi }$ 
on the cosmological constant barrier $w_{\phi }=-1$ at relatively recent times, which has 
a significant impact on the 
expected number density of massive CDM halos at different cosmic epochs. Due to this peculiar feature this type of models has been 
dubbed the ``bouncing" cDE scenario \citep[][]{Baldi_2011c}.

While in the case of the exponential potential models the scalar field always rolls down the potential with positive
velocity $\dot{\phi }>0$, reaching the normalization value $\phi = 0$ at the present time, the bouncing cDE scenario based on the SUGRA potential
is assumed
to start with the scalar field at rest in the potential minimum in the early universe, and the dynamics induced by the coupling allows
for an inversion of the direction of motion of the field during the cosmic expansion history. The peculiar dynamics of the bouncing cDE
scenario has been described in full detail by \citet{Baldi_2011c}, to which we refer for a more comprehensive treatment of the
basics of this model.
\ \\

To complete the definition of the range of models considered in the {\small CoDECS} project we need to specify
the coupling function $\beta _{c}(\phi )$, for which we will assume the exponential form proposed by \citet{Amendola_2004,Baldi_2011a}:
\begin{equation}
\beta _{c}(\phi ) \equiv \beta _{0}e^{\beta _{1}\phi }
\end{equation}
and we will consider in our analysis both the case of a constant coupling ($\beta _{1}=0$) and of an exponentially growing coupling ($\beta _{1}>0$).

All the cosmological models included in the {\small CoDECS} suite, with the relative parameters and normalizations, are summarized in Table~\ref{tab:models}.

For each of our cosmologies we compute the background evolution by numerically solving the system of coupled 
differential equations (\ref{klein_gordon}-\ref{friedmann}). For the $\Lambda $CDM and the exponential potential
models EXP001-EXP008e3 we integrate the equations backwards in time starting from the desired set of cosmological
parameters at $z=0$, by applying the integration procedure described by \citet{Baldi_2011a}. For the SUGRA cDE model, instead, since
the normalization of the field is defined at high redshifts, we integrate the system forward in time with a trial-and-error procedure until the
desired set of cosmological parameters at $z=0$ is obtained.

\begin{figure*}
\includegraphics[scale=0.45]{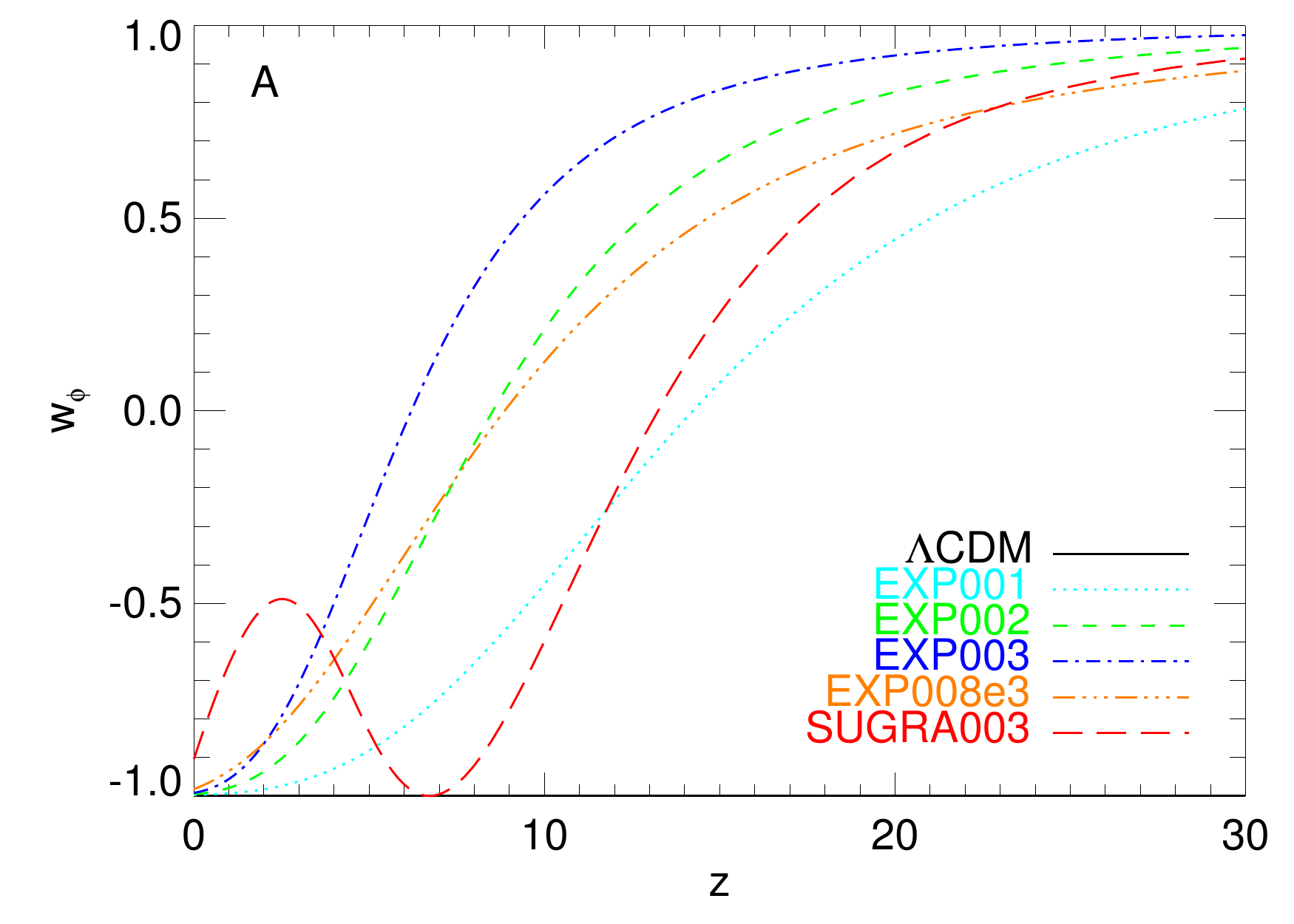}
\includegraphics[scale=0.45]{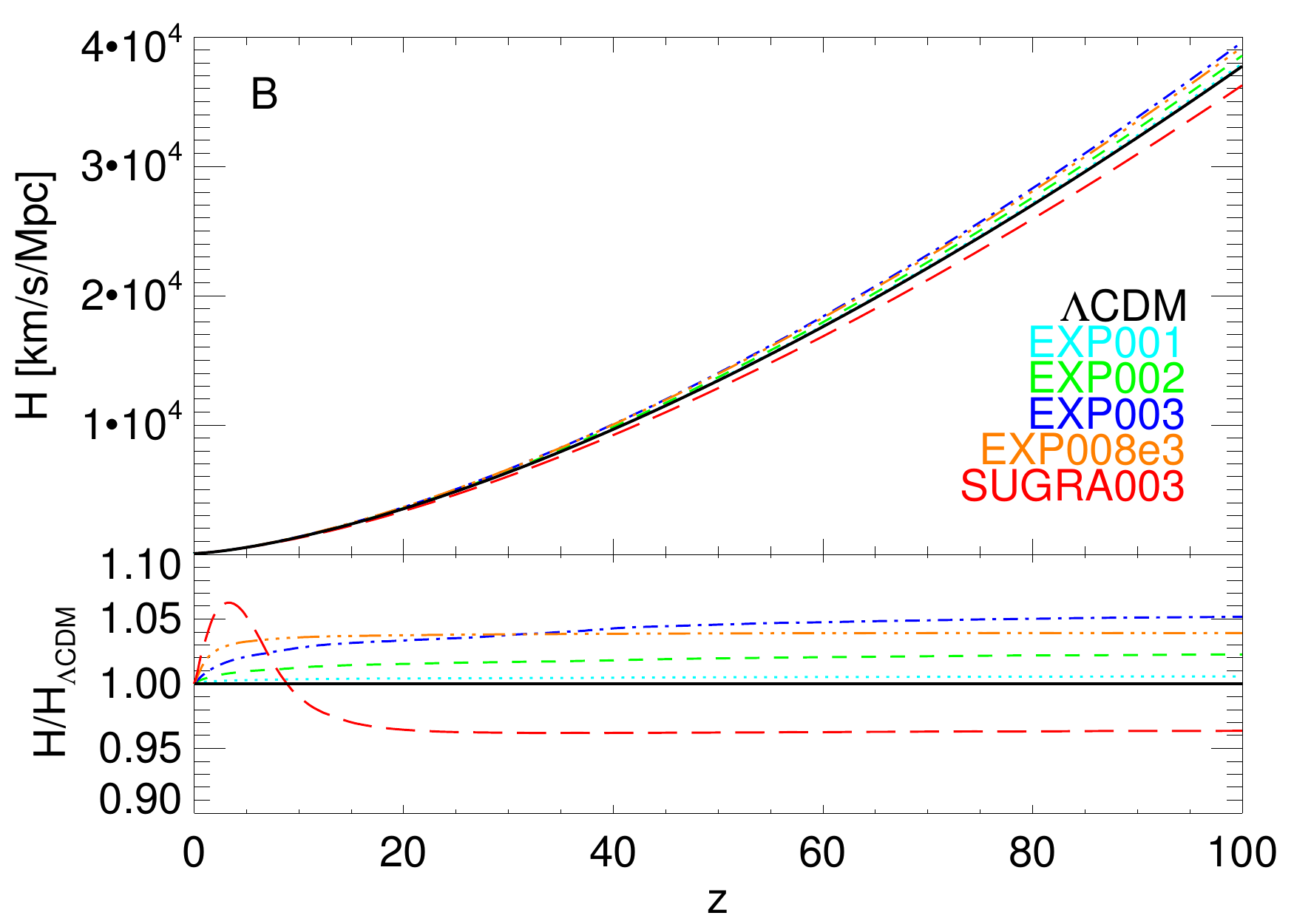}\\
\includegraphics[scale=0.45]{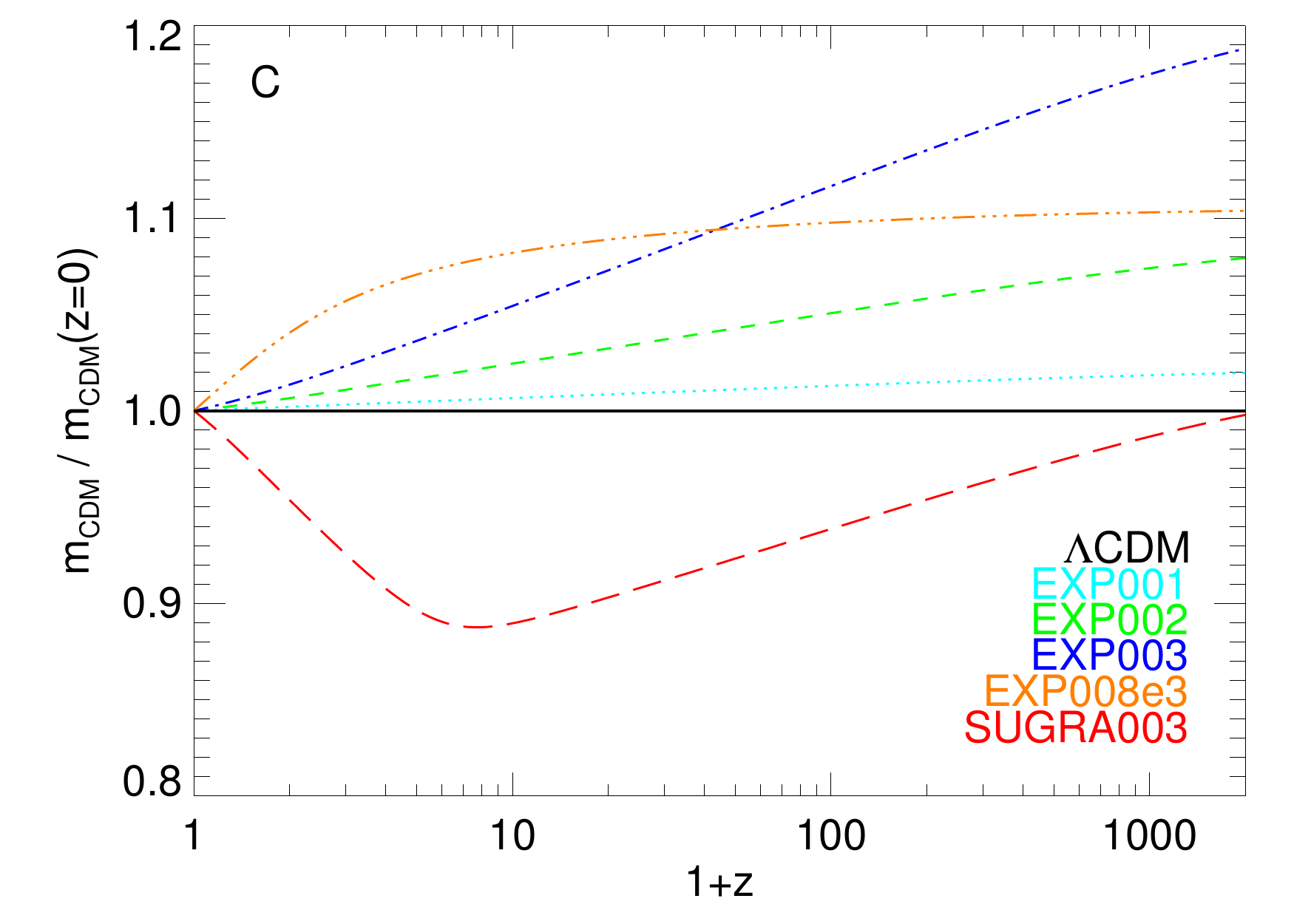}
\includegraphics[scale=0.45]{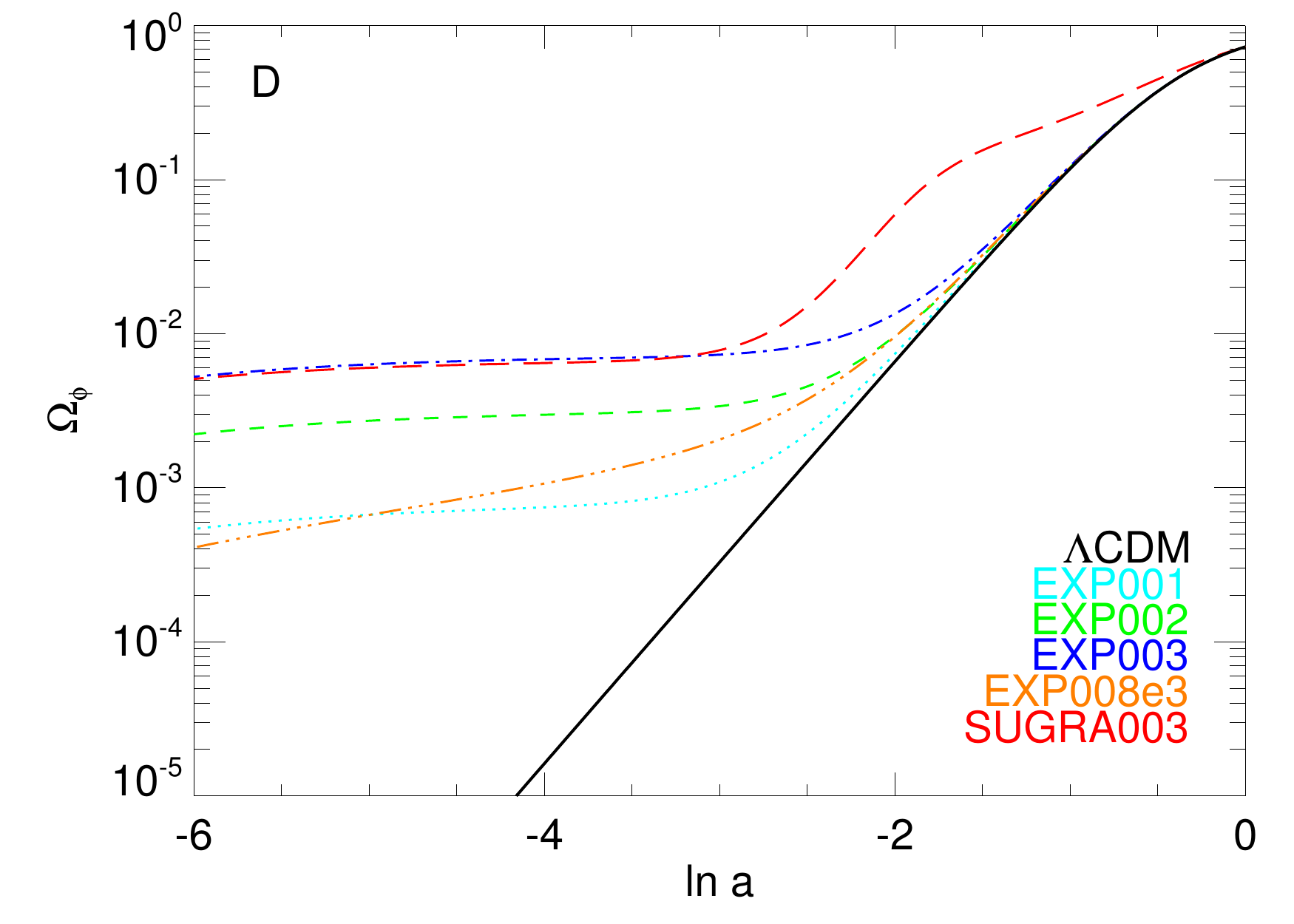}
\caption{{\em Panel A}: The equation of state parameter of DE $w_{\phi }$ as a function of redshift for all the models of the {\small CoDECS} project. The SUGRA003 model shows a ``bounce"
of the equation of state on the cosmological constant barrier $w_{\phi } = -1$ at $z_{\rm inv} \sim 6.8$ while all the other models asymptotically tend to $w_{\phi } = -1$ at $z\rightarrow 0$.
{\em Panel B}: The upper plot shows the Hubble function in km/s/Mpc for all the models under investigation, while the lower plot shows the ratio of the Hubble function
to the standard $\Lambda $CDM case. The maximum deviation is of about $6\%$ at $z\sim 5$ for the bouncing cDE model SUGRA003. {\em Panel C}: The CDM particle mass evolution
as a function of redshift. All the models feature a monotonically decreasing mass, except the bouncing cDE model SUGRA003 that shows an inversion of the mass evolution in correspondence
to the DE bounce. {\em Panel D}: The redshift evolution of the DE density parameter $\Omega _{\phi }$. The constant coupling models show a constant plateau at high z (the so called
$\phi $-MDE regime), while the growing coupling model EXP008e3 shows a slowly evolving DE fraction \citep[the so called Growing-$\phi $MDE regime, see][]{Baldi_2011a}. The 
bouncing cDE model SUGRA003 behaves like a standard constant coupling model in the early universe, but deviates from all the other models at recent epochs due to the DE bounce.}
\label{fig:background}
\end{figure*}

The evolution of the DE equation of state parameter $w_{\phi }$, of the Hubble function $H(z)$, 
of the CDM mass $m_{c}(z)/m_{c,0}$, and of the DE fractional density $\Omega _{\phi }$
are shown in Fig.~\ref{fig:background}, in panels A, B, C, and D, respectively.
\normalsize

As mentioned above, the inversion of the field motion in the SUGRA cDE model is reflected in a bounce of the equation of state $w_{\phi }$ on the so called 
cosmological constant barrier ($w_{\phi } = -1$) at $z_{\rm inv}=6.8$, as clearly shown in Fig.~\ref{fig:background}.A. As one can see from the plot, all the cDE models
have a stiff equation of state at high redshifts while they all end up at $z=0$ with values of $w_{\phi }$ that are very close to $-1$ (see also Table~\ref{tab:models}), 
with the only exception of the bouncing cDE
model that due to the bounce has a somewhat larger equation of state parameter at the present time ($w_{\phi } = -0.91$). As a consequence of its non trivial
dynamics at low redshifts, the specific bouncing cDE model considered here might therefore be challenged by present observational bounds on the DE equation of state from low-redshift
probes, as \eg Supernovae Ia. As already clearly stated by \citet{Baldi_2011c}, the SUGRA003 model should be taken as a phenomenological 
toy example of the effects that a bounce of the DE scalar field can imprint on structure formation processes, while a full parameter fitting (that goes beyond the scope
of this work) would be required in order to select the best set of parameters for a generic bouncing cDE cosmology.

The effects of the DE bounce on the background evolution of the universe are clearly visible also in Figs.~\ref{fig:background}.B and \ref{fig:background}.C, 
where the Hubble function and the CDM mass evolution are plotted, respectively, as a function of redshift. The former plot shows a clear inversion
of the deviation from the $\Lambda $CDM Hubble function in correspondence of the bounce for the SUGRA003 model;
the magnitude of the deviation is of the order of $\sim 6\%$ at low redshifts and then slowly decreases towards very high redshifts.
Such feature does not appear, as expected, in all the other cDE models based on the exponential potential of Eq.~(\ref{exponential}).
In the latter plot, instead, the effect of the bounce is shown by the inversion of the mass evolution for the SUGRA003 model, while all the other
scenarios show a monotonic decrease in time of the CDM particle mass. It is important to notice here how the CDM mass in the SUGRA model
has roughly the same value at $z_{\rm CMB}\sim 1100$ and at the present time.

Finally, Fig.~\ref{fig:background}.D shows the evolution of the DE fractional density $\Omega _{\phi }$ as a function of the e-folding time $\ln a$ for all the models
considered in this work. The early DE nature of cDE models is clearly visible in this plot, where all the different cDE cosmologies show a non negligible
DE fraction at high redshifts, which in any case never exceeds the percent level. The effect of the bounce in the SUGRA003 model is clearly visible
also in this plot, with an evolution of $\Omega _{\phi }$ that significantly deviates both from $\Lambda $CDM and from all the other exponential cDE models
at low redshifts. 

\subsection{Linear perturbations}
\label{linear}

For all the cosmological models considered in the {\small CoDECS} suite we need to compute, besides the
specific evolution of relevant background quantities like the Hubble function and the CDM mass variation, 
as described above, also the evolution of linear density perturbations in the Newtonian limit of sub-horizon scales.
The growth factor $D_{+}(z)$ of each model will be in fact necessary in order to properly set up the initial conditions
for the N-body runs described in Section~\ref{sec:sims}. Although some fitting functions for the growth factor
in the context of cDE models have been proposed in the literature \citep{DiPorto_Amendola_2008,Baldi_etal_2010,DiPorto_Amendola_Branchini_2012}
these are valid only for the standard case of a constant coupling and for monotonic self-interaction potentials, and even in these 
simplified cases have a limited range of validity in terms of baryonic fraction and coupling strength.
For the precision level required by the present work it is therefore still necessary to rely on direct numerical computations of the 
growth factor for each specific model under investigation.
Differently from most of the previous numerical works
involving cDE scenarios \citep[see \eg][]{Maccio_etal_2004,Baldi_etal_2010,Baldi_2011a,Li_Barrow_2011}, all the
{\small CoDECS} runs assume a common normalization of the amplitude of density perturbations at $z_{\rm CMB}$, 
consistent with the latest constraints on the scalar perturbations amplitude ${\cal A}_{s}$ from the WMAP7
results \citep{wmap7}. All the models included in the {\small CoDECS} suite will therefore have a normalization of linear perturbations fully
compatible with observational constraints coming from CMB data alone.

In order to compute the growth factor $D_{+}(z)$ for all our models, we need to solve the linear perturbations equations for baryon
and CDM density fluctiations in the presence of a DE coupling. These have been derived in the literature by several authors
\citep[see \eg ][]{Amendola_2004,Pettorino_Baccigalupi_2008,Baldi_2011a}, and in Fourier space and cosmic time $t$ take the form:
\begin{eqnarray}
\label{gf_c}
\ddot{\delta }_{c} &=& -2H\left[ 1 - \beta _{c}\frac{\dot{\phi }}{H\sqrt{6}}\right] \dot{\delta }_{c} + 4\pi G \left[ \rho _{b}\delta _{b} + \rho _{c}\delta _{c}\Gamma _{c}\right] \,, \\
\label{gf_b}
\ddot{\delta }_{b} &=& - 2H \dot{\delta }_{b} + 4\pi G \left[ \rho _{b}\delta _{b} + \rho _{c}\delta _{c}\right]\,,
\end{eqnarray}
where for simplicity we have omitted the $\phi $ dependence of the coupling $\beta _{c}(\phi )$.
In Eqs.~(\ref{gf_c},\ref{gf_b}) the overdensity of CDM and baryons is defined as $\delta _{c,b}\equiv \delta \rho _{c,b,}/\rho _{c,b}$.
The factor $\Gamma _{c}$ is defined as
\begin{equation}
\label{Gamma_c_massless}
\Gamma _{c} = 1 + \frac{4}{3} \beta ^{2}_{c}(\phi )
\end{equation}
and includes the effect of the fifth-force mediated by the DE scalar field for CDM density perturbations.
The second term in the first square bracket at the right-hand-side of Eq.~(\ref{gf_c}), instead, represents the so-called ``extra friction"
arising for CDM perturbations as a consequence of momentum conservation.
The validity of Eqs.~(\ref{gf_c},\ref{gf_b}) for the scales of interest in the {\small CoDECS} suite has been
already verified in the literature both for the exponential \citep{Amendola_2004,Baldi_2011a} and for the SUGRA bouncing cDE models \citep{Baldi_2011c}.
\begin{figure*}
\label{growthfactors}
\includegraphics[scale=0.45]{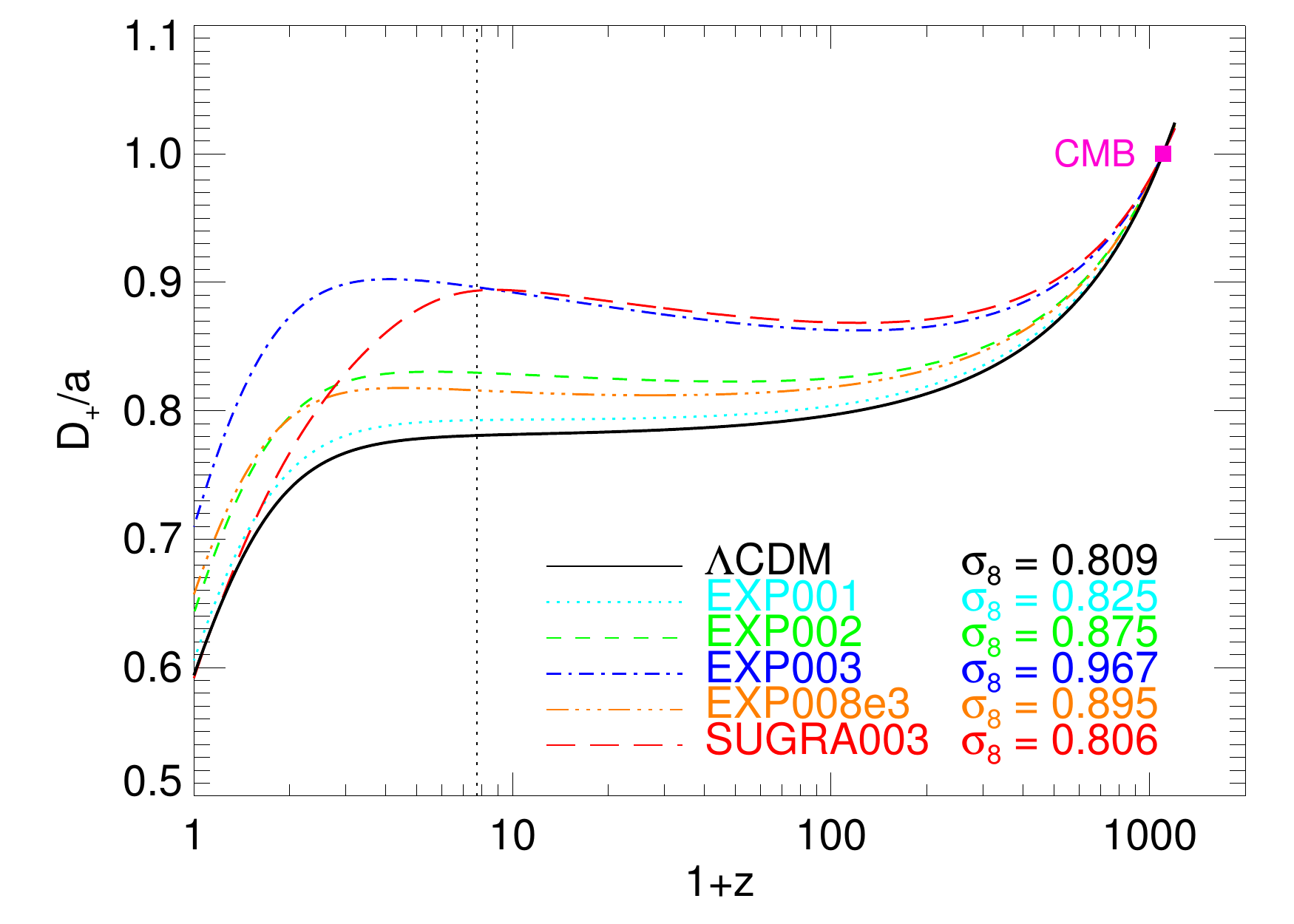}
\includegraphics[scale=0.45]{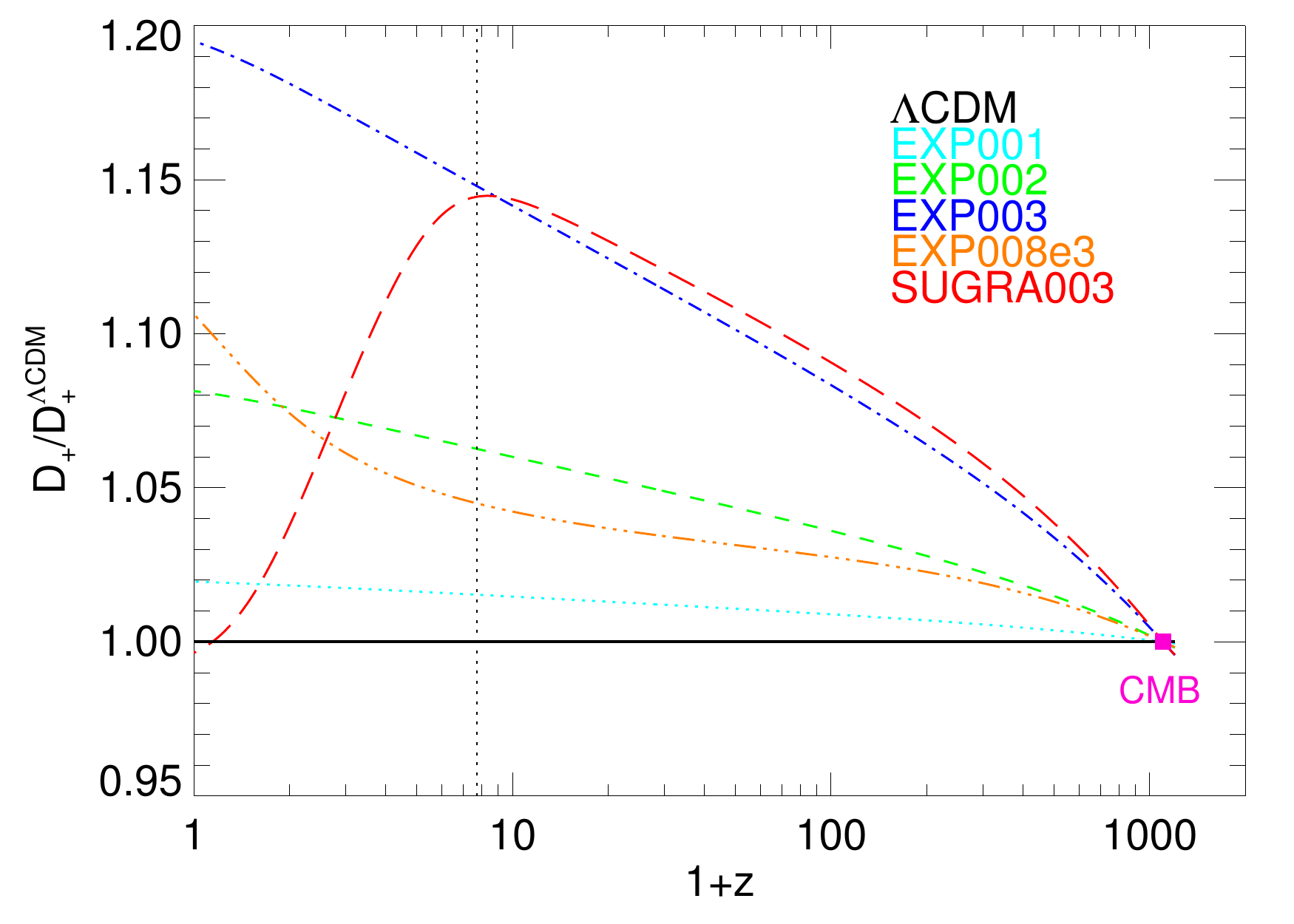}
\caption{{\em Left}: The growth factor $D_{+}(z)$ divided by the scale factor $a$ for all the models included in the {\small CoDECS} project, normalized at the redshift 
of the last scattering surface, $z_{\rm CMB}\approx 1100$. {\em Right}: The ratio of the growth factor of each cDE model to the standard $\Lambda $CDM case.
All models start with the same amplitude as $\Lambda $CDM at $z_{\rm CMB}$ but significantly deviate from the standard evolution at later times. All
the models show a monotonic increase of the deviation from $\Lambda $CDM, with the exception of the bouncing cDE model SUGRA003 that after an extended
period of strongly enhanced growth inverts the trend in correspondence to the DE bounce, and recovers the $\Lambda $CDM amplitude at $z=0$ \citep[see][]{Baldi_2011a}.}
\label{fig:linear}
\end{figure*}
\normalsize

For each of the models under investigation we therefore solve numerically Eqs.~(\ref{gf_c},\ref{gf_b}) along the corresponding background solution
assuming as a boundary condition for the integration that the ratio of baryonic to CDM perturbations at $z_{\rm CMB}$
takes the value $\delta _{b}/\delta _{c} \sim 3.0\times 10^{-3}$, as computed (for $k > 1\, h/$ Mpc) by running the publicly available 
Boltzmann code CAMB\footnote{www.camb.info} \citep[][]{camb} for a $\Lambda {\rm CDM}$ cosmology with 
the WMAP7 parameters adopted in the present study (and listed in Table~\ref{tab:parameters}). We then define the total growth factor
$D_{+}(z)$ as a combination of the baryon and CDM perturbations amplitude weighted by the relative density fraction of the 
two components, as:
\begin{equation}
D_{+}(z) \equiv \frac{1}{\Omega _{c}(z) + \Omega _{b}(z)} \left[ \Omega _{c}(z)\delta _{c}(z) + \Omega _{b}(z)\delta _{b}(z)\right] \,.
\end{equation}

The results of our integration are illustrated in Fig.~\ref{fig:linear}, where we show the evolution of the normalized growth factor $D_{+}/a$ and of
its ratio to the $\Lambda $CDM case in the left and right panels, respectively. As one can see from the figures, all the models start 
by construction with the same amplitude of density
perturbations at CMB, shown as a purple squared point in the plots, and then evolve with different growth rates until $z=0$, thereby reaching 
significantly different normalizations of the linear perturbations amplitude at the present time. This implies a different value of $\sigma _{8}$
at $z=0$ among the different cosmologies, as specified in Table~\ref{tab:models} and as clearly visible in
the figures, while all models are consistent with the same WMAP7 constraints at $z_{\rm CMB}$.

Particularly interesting are the cases of the exponential coupling model EXP008e3 (orange lines) and of the bouncing cDE model (red lines).
The former, as clearly shown in the right panel of Fig.~\ref{fig:linear}, features a mild enhancement of linear perturbations growth for most
of the expansion history of the universe, followed by a much more significant increase of the growth rate with respect to $\Lambda $CDM at low
redshifts, when the coupling exponentially grows towards its present large value $\beta _{0}=0.4$.
The latter model, instead, as discussed in full detail by \citet{Baldi_2011c}, follows a kind of opposite trend: after an extended initial phase of strongly enhanced 
growth, the evolution of density perturbations slows down dramatically as a consequence of the inversion of the scalar field motion at $z_{\rm inv}=6.8$
such that the standard $\Lambda $CDM amplitude is recovered at $z=0$. Therefore, the bouncing cDE model SUGRA003 considered in this work 
has the same normalization of linear density perturbations as $\Lambda $CDM both at $z_{\rm CMB}$ and at $z=0$, while featuring a peak in its
deviation from $\Lambda $CDM around $z\sim 10$.
This peculiar evolution is clearly visible by eye in both panels of Fig.~\ref{fig:linear}
just by following the different paths of the black ($\Lambda $CDM) and red (SUGRA003) lines, and is expected to have 
significant consequences on structure formation at different cosmic epochs \citep[][]{Baldi_2011c}.

\section{The CoDECS simulations suite}
\label{sec:sims}

As already mentioned above, the main purpose of the present paper is to present and discuss a vast numerical program for interacting DE cosmologies
that has been recently carried out by the author and that is now made available to the scientific community for public use.
We refer to this numerical enterprise as the COupled Dark Energy Cosmological Simulations ({\small CoDECS})
project, and in the present Section we will discuss the main features of the numerical implementation adopted for the different sets of runs 
carried out in the context of this project.
\begin{figure*}
\includegraphics[scale=0.5]{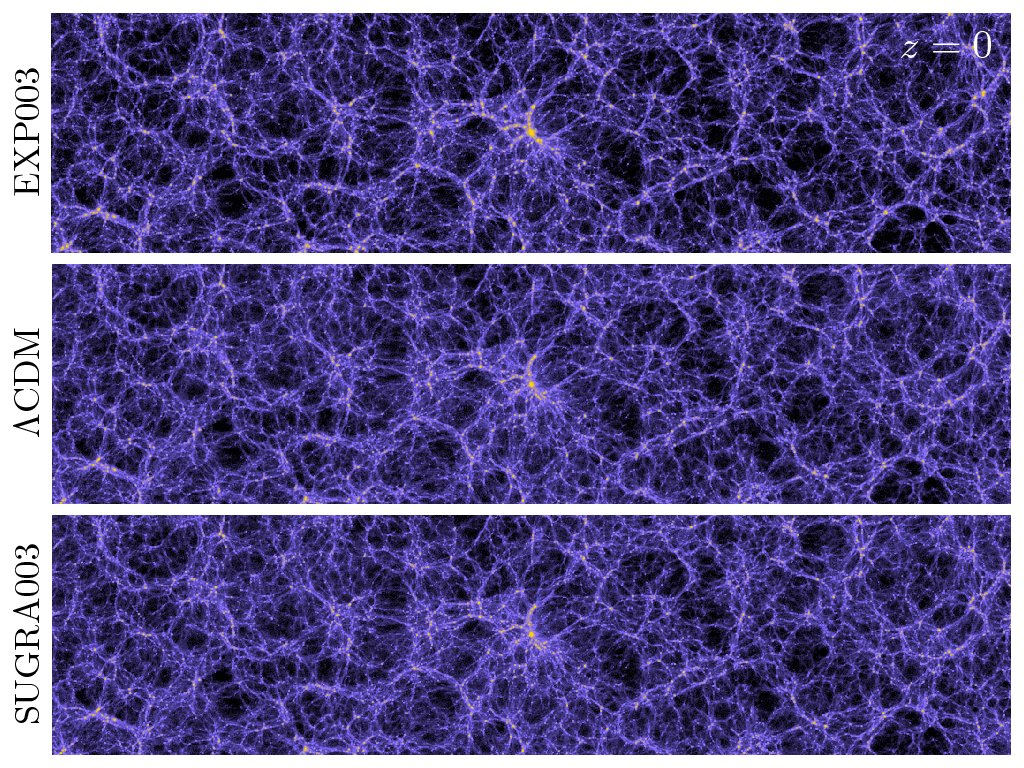}
\caption{The CDM density distribution in a slice with size $1000\times 250$ Mpc$/h$ and thickness $30$ Mpc$/h$ as extracted from the {\small L-CoDECS}
simulations of a few selected models. The middle slice shows the case of the standard $\Lambda $CDM cosmology, while the top slice is taken from the EXP003 simulation
and the bottom slice from the bouncing cDE model SUGRA003. While the latter model shows basically no difference with respect to $\Lambda $CDM at $z=0$, due to the very similar value of $\sigma _{8}$
for the two models, clear differences in the overall density contrast and in the distribution of individual halos can be identified by eye for the EXP003 cosmology. (A higher resolution version of this figure
is available online through the {\small CoDECS} website, see the Appendix)}
\label{fig:slice}
\end{figure*}
\normalsize

\begin{figure*}
\includegraphics[scale=0.5]{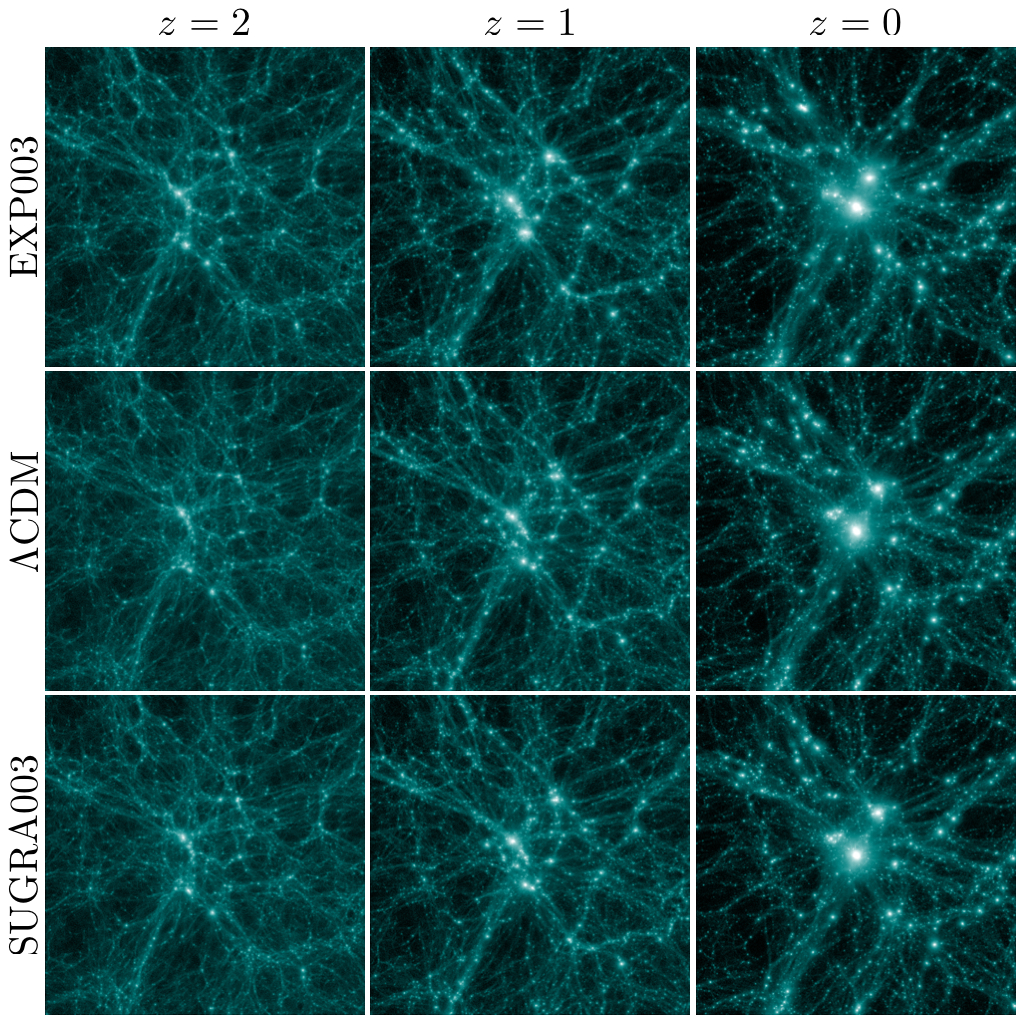}
\caption{The gas density distribution during the formation process of a massive galaxy cluster as extracted from the {\small H-CoDECS} runs for the same three
models shown in Fig.~\ref{fig:slice}. Also in this case, differences in the overall density contrast and in the distribution of individual lumps are visible by eye
when comparing the standard $\Lambda $CDM cosmology and the EXP003 cDE model at $z=0$. However, in this case the redshift evolution shown in the figure allows to identify differences
also between $\Lambda $CDM and the bouncing cDE model SUGRA003 at higher redshifts, where the latter model appears more evolved and shows a more pronounced density contrast
as compared to the standard cosmology. (A higher resolution version of this figure
is available online through the {\small CoDECS} website, see the Appendix)}
\label{fig:halo}
\end{figure*}
\normalsize

\subsection{Initial Conditions}

Initial conditions for N-body simulations are generated by displacing particles from
a homogeneous distribution in order to set up a random-phase realization of the linear matter power spectrum of the cosmological model
under consideration according to Zel'dovich approximation \citep{Zeldovich_1970}. The particles displacements are then rescaled to the
desired amplitude of density perturbation at some high redshift $z_{i}$ when all perturbation modes included in the simulation box
are still evolving linearly. This redshift is then taken as the starting redshift of the simulation, and the corresponding particle distribution as
the initial conditions for the N-body run.

For all the simulations considered in the {\small CoDECS} project we generate initial conditions at $z_{i} = 99$ by displacing particles
from a homogeneous {\em glass} distribution \citep{White_1994,Baugh_etal_1995}. The initial power spectrum corresponds to the one obtained from CAMB 
at $z=99$ for a $\Lambda $CDM cosmology with the 
cosmological parameters of Table~\ref{tab:parameters}, and includes baryonic acoustic oscillations (BAO). The same power spectrum is used
for both CDM and baryon particles, which implies discarding possible effects due to a residual offset between the amplitude of perturbations in the
two fluids at $z=99$ \citep[see \eg][]{Naoz_Yoshida_Barkana_2010}, and for all the different cosmologies under consideration, thereby not considering
possible distortions of the transfer function due to early effects of the coupling. This choice is made in order to allow for a direct comparison of the
impact that cDE cosmologies have on the growth of structures between $z_{\rm CMB}$ and the present. Furthermore, the discarded effects are expected 
to have a much weaker impact on structure formation processes as compared to the DE interactions under investigation \citep[][]{Baldi_etal_2010}.
The particles displacements obtained for each simulation as a random-phase realization of the same initial power spectrum 
are then rescaled to the same amplitude at $z_{\rm CMB}$,
corresponding to the WMAP7 normalization of scalar perturbations ${\cal A}_{s} = 2.42\times 10^{-9}$. 
As we also employ the very same random seed for all the simulations, 
the resulting particle displacements are exactly the same for all the cosmological models at
$z_{\rm CMB}$ and the initial conditions for each individual run are then generated by rescaling 
forward the displacements amplitudes to $z_{i}=99$ with the correct growth factor $D_{+}(z)$ (see Fig.~\ref{growthfactors}) computed for each cosmological model as described in Sec.~\ref{models}.

Once particles positions have been assigned, velocities are computed in Fourier space 
according to the linear perturbation theory relation $v(k,a) \propto f(a)\delta (k,a) $,
where the growth rate function $f(a)\equiv d\ln D_{+}/d\ln a$ can be derived for each model from the corresponding computed growth factor. 

\subsection{The N-body Simulations}

The {\small CoDECS} suite includes at the present time the six different cosmological models listed in Table~\ref{tab:models} that have been 
extensively described in Sec.~\ref{models}. For all these models we have run so far two different sets of cosmological N-body simulations, 
named {\small L-CoDECS} and {\small H-CoDECS}. Both sets of simulations consist of a cosmological volume with periodic boundary conditions
filled with an equal number of CDM and baryonic particles, but differ from each other for the scale and for the physical processes included
in the runs. All simulations have been carried out with the modified version by \citet{Baldi_etal_2010} of the widely used parallel Tree-PM
N-body code {\small GADGET} \citep{gadget-2}, specifically developed to include all the additional physical effects that characterize 
cDE models \citep[see][for a detailed description of the implementation
of cDE in the code]{Baldi_etal_2010}.
\ \\

The {\small L-CoDECS} simulations have a box size of $1$ comoving Gpc$/h$ aside and include $1024^{3}$ CDM and baryon
particles for a total particle number of $2\times 1024^{3}\approx 2\times 10^{9}$. The mass resolution at $z=0$ for this
set of simulations is $m_{c}=5.84\times 10^{10}$ M$_{\odot }/h$ for CDM and $m_{b}=1.17\times 10^{10}$ M$_{\odot }/h$ for
baryons. Despite the presence of baryonic particles these simulations do not include hydrodynamics and are therefore purely collisionless
N-body runs. The inclusion of baryonic particles is necessary in order to realistically follow the growth of structures in the context
of specific cosmological scenarios (as the cDE models under discussion here) where baryons and CDM do not obey the same dynamical equations. In other words, if the 
fraction of uncoupled matter given by the baryons was omitted, the effect of the coupling on structure formation would be significantly
overestimated in the simulations leading to largely incorrect results. On the other hand, the different dynamics followed by CDM and baryonic particles
imprints characteristic features on the relative distribution of the two fluids which might have interesting observational implications,
and therefore the inclusion of collisionless baryonic particles in the runs offers also the possibility to estimate such effects, as we will discuss in detail in Sec.~\ref{bias}.

In Fig.~\ref{fig:slice} we show the CDM density distribution at $z=0$ in a slice of $1000\times 250$ Mpc$/h$ for the standard $\Lambda $CDM cosmology
(middle) and for two of our cDE models, namely the most extreme model with exponential potential and constant coupling EXP003 (top)
and the bouncing cDE model SUGRA003 (bottom). The slice is $30$ Mpc$/h$ thick and is centered on the most massive halo identified for the 
$\Lambda $CDM run. Although the main pattern of the large-scale cosmic web looks very similar in the three models (as a consequence of the identical random phases
adopted for the generation of the initial conditions in each simulation), some differences can be identified even by eye when comparing the density contrast
of the most concentrated objects as well as of the largest voids. The EXP003 model appears clearly more structured and with a more pronounced density contrast between 
halos (in yellow) and void regions (in black) as compared to $\Lambda $CDM. 
Furthermore, by carefully looking at the central part of the two slices, it is also possible to notice how some of the satellite halos that are falling along the filaments
onto the main central structure in the $\Lambda $CDM cosmology have already merged into bigger substructures in the EXP003 case. The SUGRA003 model shows much 
weaker signatures with respect to $\Lambda $CDM at $z=0$. This is expected since the model has the same $\sigma _{8}$ normalization
as $\Lambda $CDM at the present time. A more quantitative estimate of the differences in the abundance and distribution of massive halos is provided by the Halo Mass Function extracted from the {\small L-CoDECS} runs that is briefly sketched in Sec.~\ref{sec:HMF} and in Figure~\ref{fig:HMF}, and discussed in much finer details in \cite{Baldi_2011c} and \cite{Cui_Baldi_Borgani_2012}.
\ \\

The {\small H-CoDECS} simulations are instead adiabatic hydrodynamical simulations in a cosmological box of $80$ Mpc$/h$ aside
filled with $512^{3}$ CDM and gas particles, where the gas particles represent the uncoupled baryonic fraction.
The mass resolution at $z=0$ for this set of simulations is $m_{c}=2.39\times 10^{8}$ M$_{\odot }/h$ for CDM and $m_{b}=4.78\times 10^{7}$ M$_{\odot }/h$ for
baryons.
Adiabatic hydrodynamical forces on the gas particles are computed by means of the entropy conserving formulation of
{\em Smoothed Particle Hydrodynamics} \citep[SPH,][]{Springel_Hernquist_2002} and other radiative processes such as 
gas cooling, star formation, or feedback mechanisms are not included in the simulations.

In Fig.~\ref{fig:halo} we show the formation of a massive galaxy cluster in the same three cosmological models displayed in Fig.~\ref{fig:slice} by plotting the
gas density contrast in a slice of $40\times 40$ Mpc$/h$ centered in the position of the most massive halo identified at $z=0$ in the $\Lambda $CDM  
simulation of the {\small H-CoDECS} suite, for three different redshifts $z=0$, $1$, and $2$.
In this case, since the figure shows the redshift evolution of the three models, it is possible to identify the differences that characterize the growth of structures in these 
selected scenarios. Also this comparison shows the same features already described above for the comparison between $\Lambda $CDM and EXP003. By comparing the
two cosmologies at $z=0$, in fact, it clearly appears how the latter has a more evolved structure of the cosmic web around the central massive halo and shows a larger density contrast between the highly nonlinear overdense regions and the voids.
It also appears much more clearly than before, from these plots, the lack of small structures in the EXP003 universe with respect to $\Lambda $CDM, due to the faster merging processes occurring in the cDE
cosmology. 
Despite this, the relative abundance of substructures as a function of their fractional mass with respect to their host main halo is not significantly affected by the coupling, as will appear from the Subhalo Mass Function displayed in Fig.~\ref{fig:subHMF} below.

The higher amplitude of density perturbations in the EXP003 model is visible also at higher redshifts, especially at $z=1$, 
where the difference in the structure of the cosmic web
around the central halo between EXP003 and $\Lambda $CDM is particularly evident, as well as the different merging history of the two models.
Particularly interesting is in this case also the comparison between $\Lambda $CDM and the bouncing cDE model SUGRA003. At $z=0$,
as already discussed above, the two models show much weaker differences
with respect to the case of the EXP003 cosmology, due to the comparable amplitude of their linear density perturbations. However, at high redshifts (\eg at $z=2$)
the SUGRA003 model appears clearly more evolved as compared to $\Lambda $CDM: the main filamentary structure and the high density peaks appear more luminous
than the corresponding $\Lambda $CDM case, and the main halos are more extended due to the merging of the surrounding satellite structures. This shows the main
characteristic feature of the bouncing cDE scenario: a larger amplitude of density perturbations (and a correspondingly higher number of massive clusters)
with respect to $\Lambda $CDM at high redshifts, 
followed by a reconnection of the two models that show very little differences in their large-scale structures at the present time \citep[][]{Baldi_2011c}.

High-resolution versions of the images displayed in Figs.~\ref{fig:slice}~and~\ref{fig:halo} are available online through the {\small CoDECS} website (see the Appendix). 

\section{Basic Results}
\label{sec:results}

We describe here some basic results arising from the analysis of the {\small CoDECS} simulations, focusing on the nonlinear
matter power spectrum, on the effects of the coupling on the relative evolution of CDM and baryon density perturbations,
and on the halo and subhalo mass functions.
As the simulations outputs are now publicly available, more detailed and extended analysis might be carried out in the near future
also by other independent groups. This is the main aim of the {\small CoDECS} public release discussed in the Appendix to this paper.

\subsection{The nonlinear matter power spectrum}
\begin{figure*}
\includegraphics[scale=0.35]{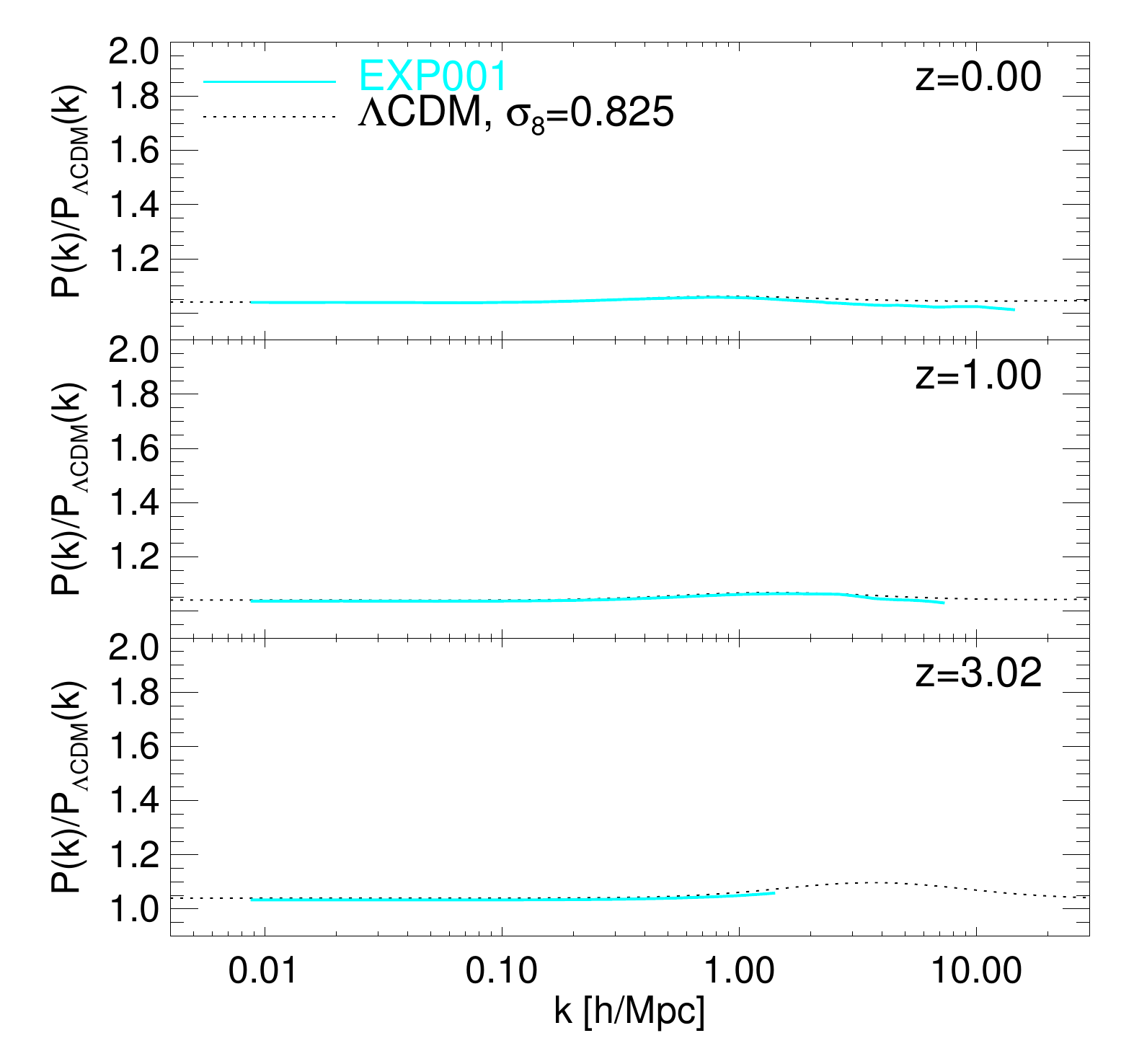}
\includegraphics[scale=0.35]{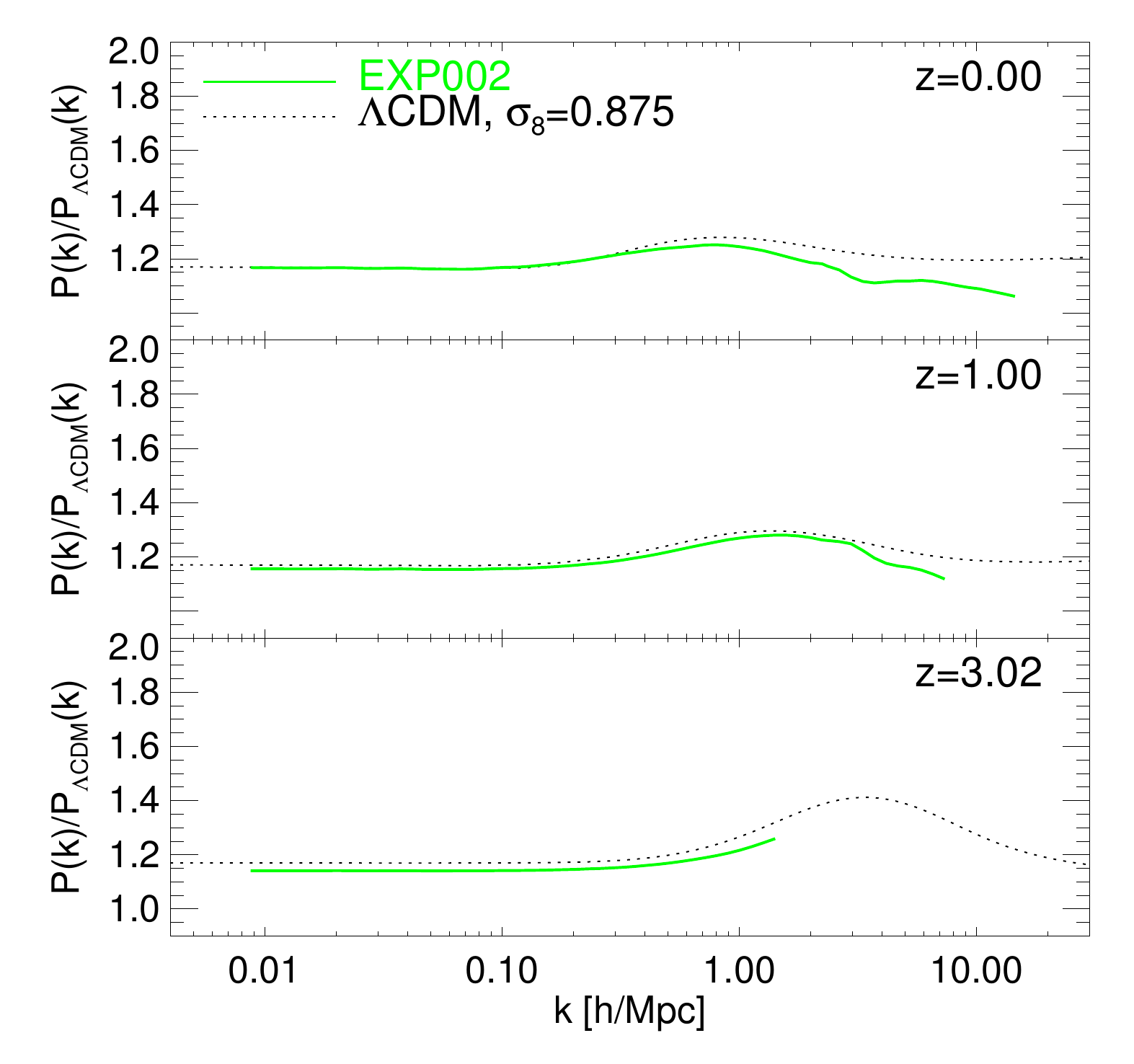}
\includegraphics[scale=0.35]{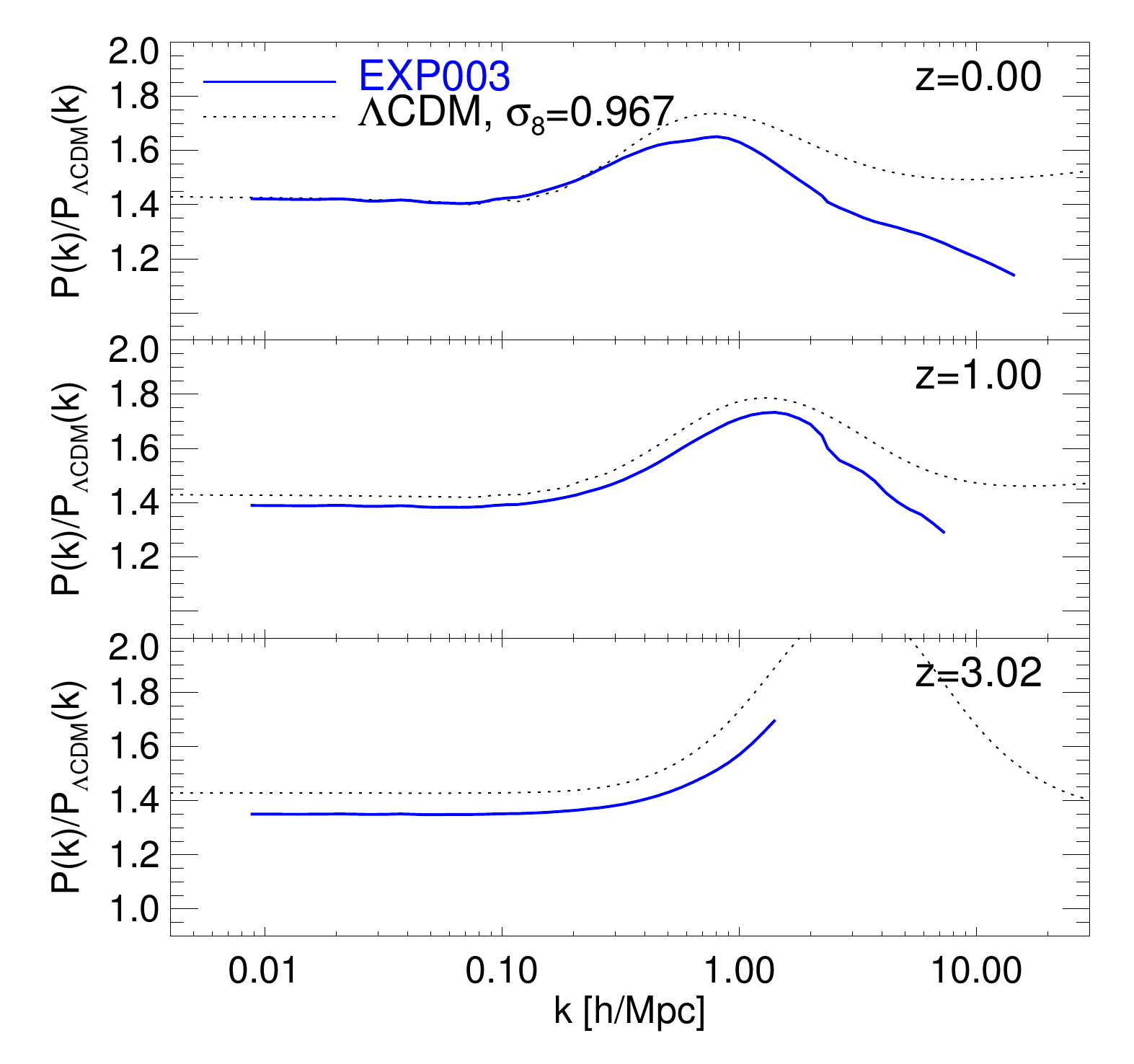}\\
\includegraphics[scale=0.35]{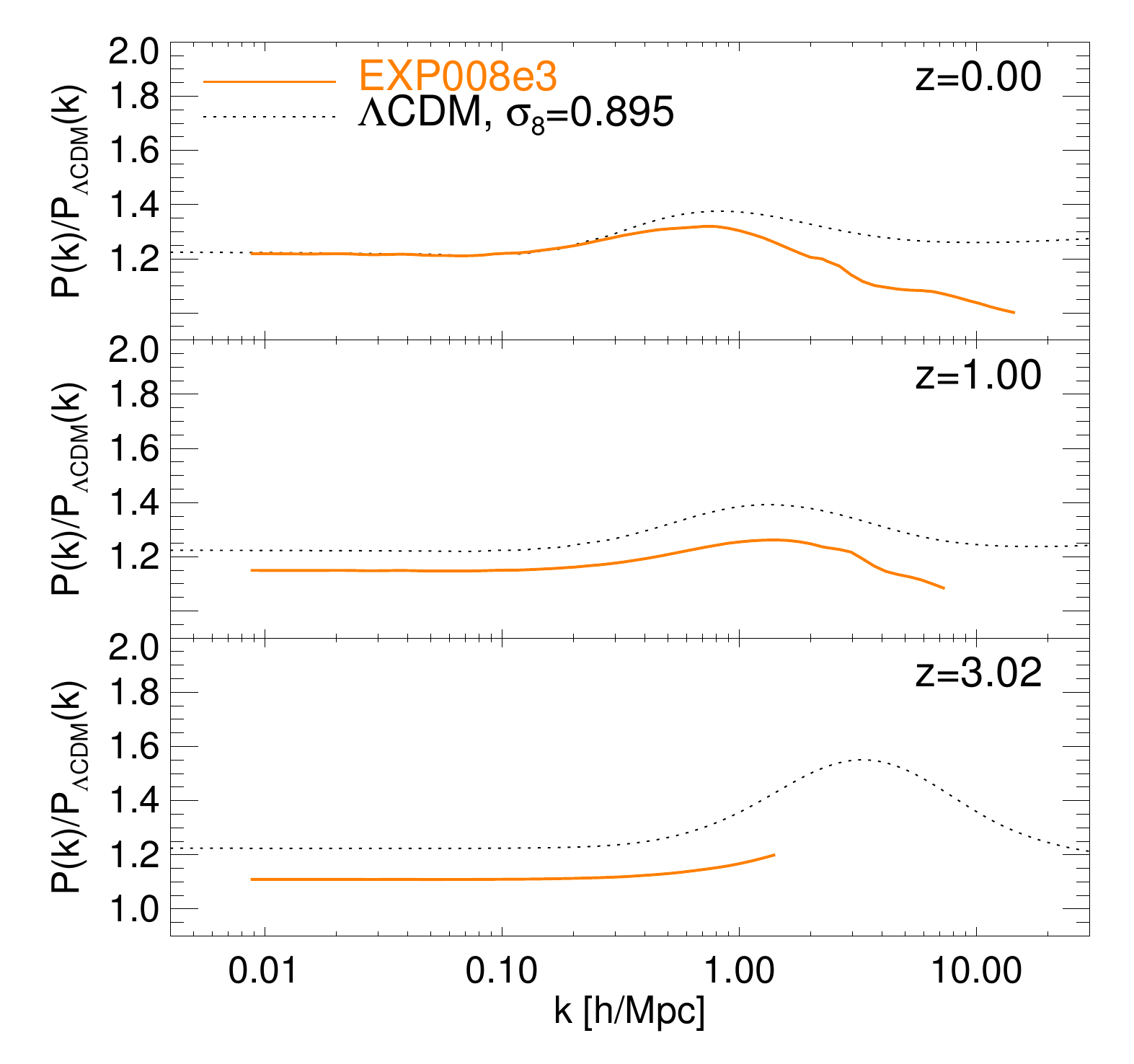}
\includegraphics[scale=0.35]{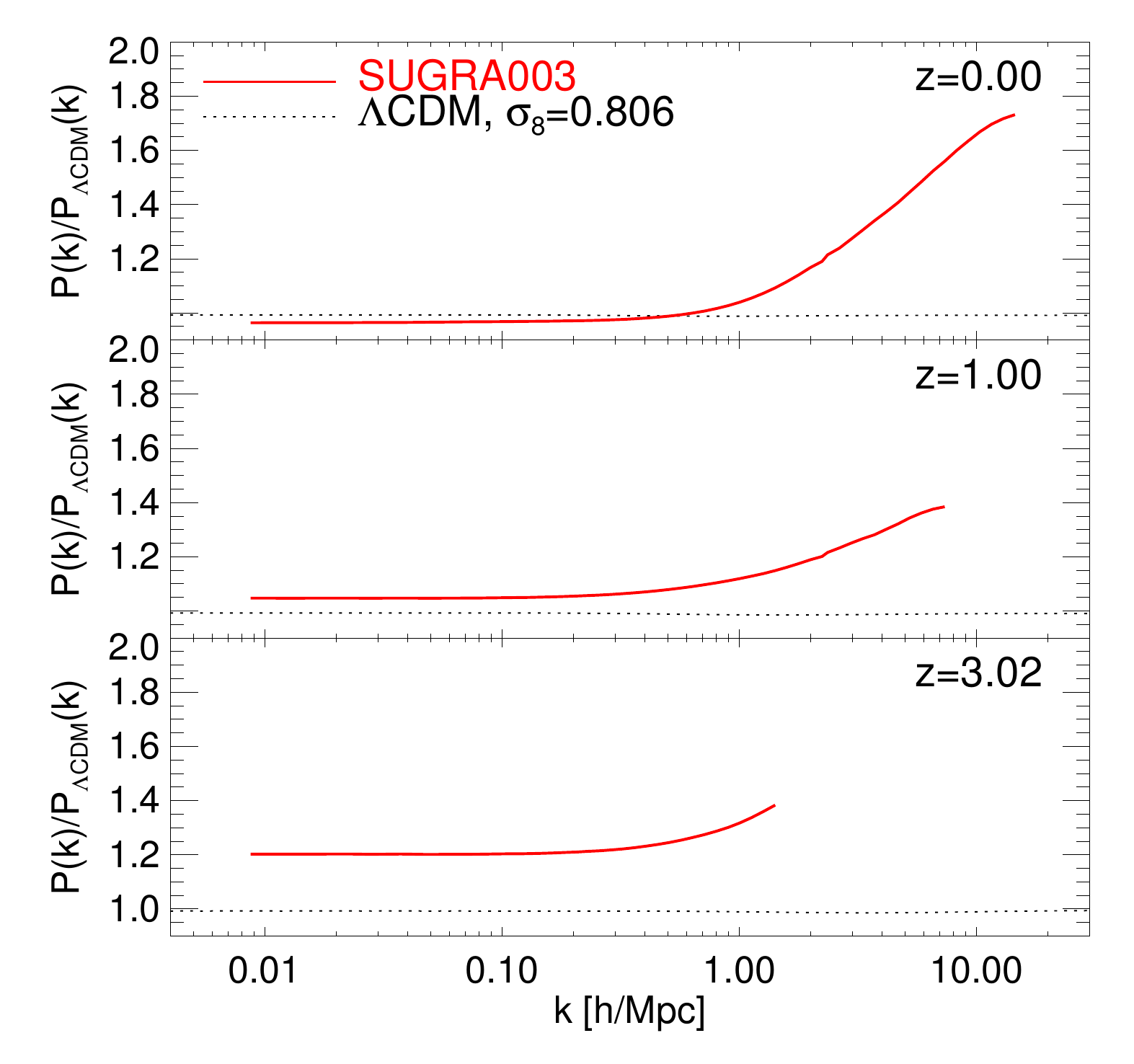}
\caption{The ratio of the nonlinear matter power spectrum to the standard $\Lambda $CDM cosmology for all the cDE models under discussion,
as extracted from the {\small L-CoDECS} runs (solid), and the ratio of the nonlinear matter power spectrum for a $\Lambda $CDM model with the same 
$\sigma _{8}$ as the respective cDE model and the fiducial $\Lambda $CDM cosmology with $\sigma _{8}=0.809$, as computed by means of the 
{\small HALOFIT} routine of the public code {\small CAMB} (dotted). Although at linear scales the two quantities are completely degenerate at $z=0$, 
the different redshift evolution of the linear power spectrum amplitude allows to break the degeneracy. Furthermore, even at $z=0$, the nonlinear regime shows 
a clear difference between the effects of cDE and of a high value of $\sigma _{8}$ due to the distortion determined by the DE coupling on the matter power spectrum,
be it either a suppression (as for constant and variable coupling models) or an enhancement (as for bouncing cDE).}
\label{fig:powerspec}
\end{figure*}
\normalsize

For the {\small L-CoDECS} runs we compute the density power spectrum of the different components $P_{b}(k)$,
$P_{c}(k)$, $P_{\rm M}(k)$ (for the CDM, baryon, and the total matter components, respectively) by determining the individual density fields 
on a grid with the same size of the PM grid (\ie $1024^{3}$ nodes) through a Cloud-in-Cell mass assignment. 
This procedure provides a determination of the nonlinear matter power spectrum 
up to the Nyquist frequency of the PM grid, which for the {\small L-CoDECS} runs corresponds to $k_{\rm Ny} = \pi N/L \approx 3.2\, h/$Mpc.
Beyond this limiting wave number we estimate the power spectrum by means of the folding method of \citet{Jenkins_etal_1998,Colombi_etal_2009},
and we smoothly interpolate the two estimations around $k_{\rm Ny}$ in order to obtain a single determination of the matter power spectrum at different redshifts
for all the simulations of the sample. The small scale power spectrum is then truncated at the frequency where the shot noise reaches $10\%$ of the measured power.

With the nonlinear power spectra determined as discussed above, we can investigate the effects that the coupling induces on the growth of structures
and highlight how differently these effects characterize the linear and the nonlinear regimes of structure formation. We should stress once more that the {\small CoDECS}
runs differ from all the previous N-body simulations of interacting DE models as they feature a common normalization of density perturbations at CMB, while 
previous studies \citep[][]{Maccio_etal_2004,Baldi_etal_2010,Li_Barrow_2011,Baldi_2011a} have almost always assumed a common normalization of linear perturbations at $z=0$. Our results therefore represent the first realistic estimation
of the effects of cDE models for a growth history of density perturbations fully compatible with CMB data. 

As a consequence of the common normalization of perturbations at CMB
and of the different growth factors determined by the modified expansion history and by the fifth-force and extra friction terms in Eq.~\ref{gf_c}, 
the various cDE models will have in general a different amplitude of density perturbations at
any redshift $z < z_{\rm CMB}$ and in particular at the present time $z=0$. This will imply a different value of $\sigma _{8}$ from model to model, as
summarized in Table~\ref{tab:models}, with the only exception of the SUGRA003 bouncing cDE model that has been constructed in order to have the same linear
normalization as $\Lambda $CDM both at $z_{\rm CMB}$ and at the present epoch. 

Furthermore, the coupling between DE and CDM has very 
different effects on the growth of structures in the linear and in the nonlinear regimes. In particular,
as it has been shown in previous studies \citep[see \eg ][for a detailed discussion of this effect]{Baldi_2011b}, the extra friction term of Eq.~\ref{gf_c}
for the case of standard cDE models (as the exponential models with constant coupling discussed in this work)
accelerates CDM particles along their direction of motion, and therefore
acts in the direction of enhancing the growth of structures in the linear regime, while in the nonlinear regime, due to the onset of virialization processes and the 
appearance of non-radial velocities in the local peculiar flow of particles, it slows down the contraction of collapsed structures thereby slowing
down structure formation processes. This determines a reduction of the small scale nonlinear power in cDE models 
with respect to the expected power based on their large-scale linear normalization. Such effect is completely inverted for the case of the SUGRA bouncing cDE
model, where the sign of the extra-friction term changes in correspondence to the ``bounce" of the DE scalar field at $z_{\rm inv}\sim 6.8$.
For this model, in fact, since $\dot{\phi }\beta_{c}(\phi ) < 0$ for $z < z_{\rm inv} \approx 6.8$, the friction term acts as a proper 
friction during the late stages of structure formation, 
thereby slowing down particles along the direction of their motion. This implies that the friction term alleviates the effect of the fifth-force by slightly
slowing down the growth of structures in the linear regime at low redshifts, while it favors the contraction of collapsed objects in the nonlinear regime.

This opposite behavior of the extra-friction term in the linear and nonlinear regimes determines specific types of distortion of the matter power
spectrum that allow to break the degeneracy of the effect of the coupling at the linear level with other cosmological parameters -- as \eg $\sigma _{8}$ -- as shown in Fig.~\ref{fig:powerspec}.
In each of the plots of Fig.~\ref{fig:powerspec}, in fact, we display with solid lines the ratio of the nonlinear matter power spectrum of each cDE model to the fiducial
$\Lambda $CDM cosmology with $\sigma _{8}=0.809$ as extracted from the {\small L-CoDECS} simulations. 
As a useful comparison, the dotted line shows the ratio of the nonlinear matter power spectra computed with
the Boltzmann code {\small CAMB} by means of the {\small HALOFIT} fitting routine \citep[][]{Smith_etal_2003}
for a $\Lambda $CDM model with a $\sigma _{8}$ normalization corresponding to the one of
the respective cDE model and for the fiducial $\Lambda $CDM cosmology. In other words, the solid lines show the ratio $P_{\rm sim}({\rm cDE})/P_{\rm sim}(\Lambda {\rm CDM})$
while the dotted lines show the ratio $P_{{\rm CAMB}}({\rm cDE}\, \sigma _{8})/P_{{\rm CAMB}}({\rm fiducial}\, \sigma _{8})$.
\ \\

The three upper plots of Fig.~\ref{fig:powerspec} display these ratios for the three standard cDE models EXP001, EXP002, and EXP003 at different redshifts. 
For very small couplings (as \eg for EXP001, cyan line) the effect on the matter power spectrum at $z=0$ appears to be fully degenerate with $\sigma _{8}$, as the two ratios
look indistinguishable at all scales. However, for larger couplings (EXP002, green line) a suppression of power in the cDE model with respect to a 
$\Lambda $CDM realization with identical $\sigma _{8}$ is clearly visible at small scales. The effect becomes more prominent for the strongest coupling
(EXP003, blue lines), where a significant deviation between the cDE model and a $\Lambda $CDM cosmology with $\sigma _{8}=0.967$ appears at $z=0$
for $k>0.3\, h/$Mpc. Therefore, even at $z=0$ the nonlinear scales of the matter power spectrum allow to distinguish between a cDE scenario and a $\Lambda $CDM
model with the same large-scale amplitude of linear perturbations, removing the apparent degeneracy that arises when only linear scales are considered.
This difference might be beyond the discriminating power of present observational probes for very small couplings, but represents in any case a distinctive 
signature of cDE scenarios.

Furthermore, by looking at the redshift evolution of the large-scale matter power in the three upper plots of Fig.~\ref{fig:powerspec} it is clearly possible
to distinguish, at least in principle, a cDE model from a $\Lambda $CDM cosmology with high $\sigma _{8}$ by the different evolution of the amplitude of density perturbations associated to the different linear growth factors displayed in Fig.~\ref{fig:linear}. All the standard cDE models EXP001-EXP003, in fact, show a lower amplitude of the linear matter power at high redshift as compared to what
expected for a $\Lambda $CDM cosmology with the same value of $\sigma _{8}$ at $z=0$.
\ \\

The two lower plots of Fig.~\ref{fig:powerspec} show the same quantities for the exponential coupling model EXP008e3 and for the bouncing cDE model SUGRA003.
Interestingly, while the former shows a similar behavior as the standard cDE models with constant coupling, 
with a large-scale amplitude at $z=0$ fully consistent with the linear growth expected for this model, and with a 
small-scale suppression of power due to the nonlinear effect of the friction term, the latter model is found to have the opposite trend both concerning the redshift dependence of the large-scale 
perturbations amplitude and the small-scale distortion of the nonlinear matter power. In particular, the linear amplitude of density perturbations is fully consistent with the fiducial 
$\Lambda $CDM cosmology at the present time, while a significantly higher normalization with respect to the fiducial model clearly appears at high redshifts. This offers a direct way to constrain this scenario and to account for a possible anomalous clustering at high z \citep[][]{Baldi_2011c} with the next generation of 
large galaxy surveys. At the nonlinear level, instead, the model
shows a strong increase of the small scale power at low redshifts in a range of scales ($k \sim 1-10 \, h/$Mpc) potentially accessible to accurate weak lensing measurements. 

Besides its peculiar impact on the redshift evolution of the cluster mass function \citep[see][for a study of the expected cluster number counts]{Baldi_2011c}, that remains the main motivation behind the whole bouncing cDE scenario, 
these specific signatures identified for the first time by the present study 
clearly provide a handle to observationally test this class of cosmologies. 

\subsection{The gravitational bias}
\label{bias}
\begin{figure}
\includegraphics[scale=0.45]{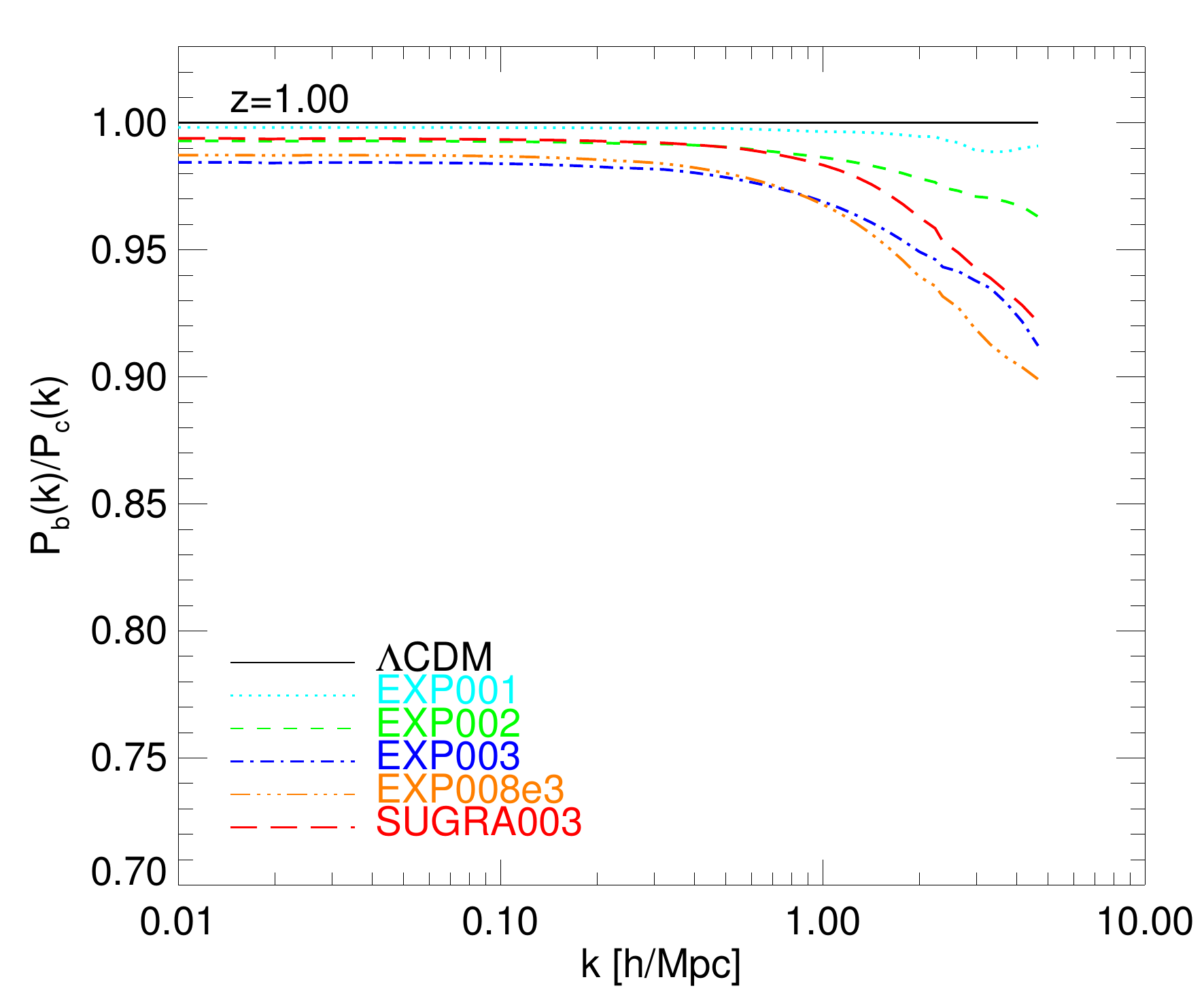}\\
\includegraphics[scale=0.45]{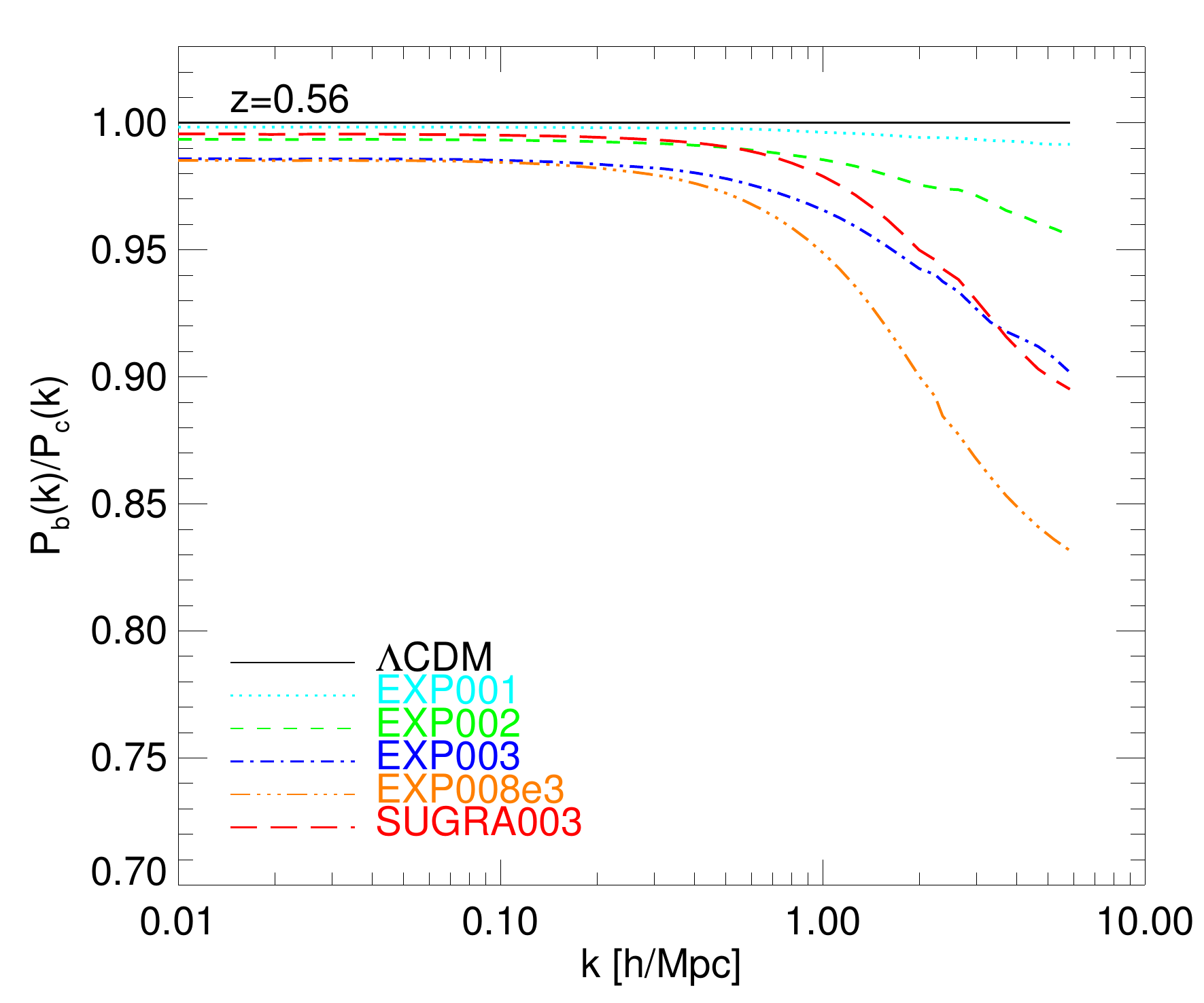}\\
\includegraphics[scale=0.45]{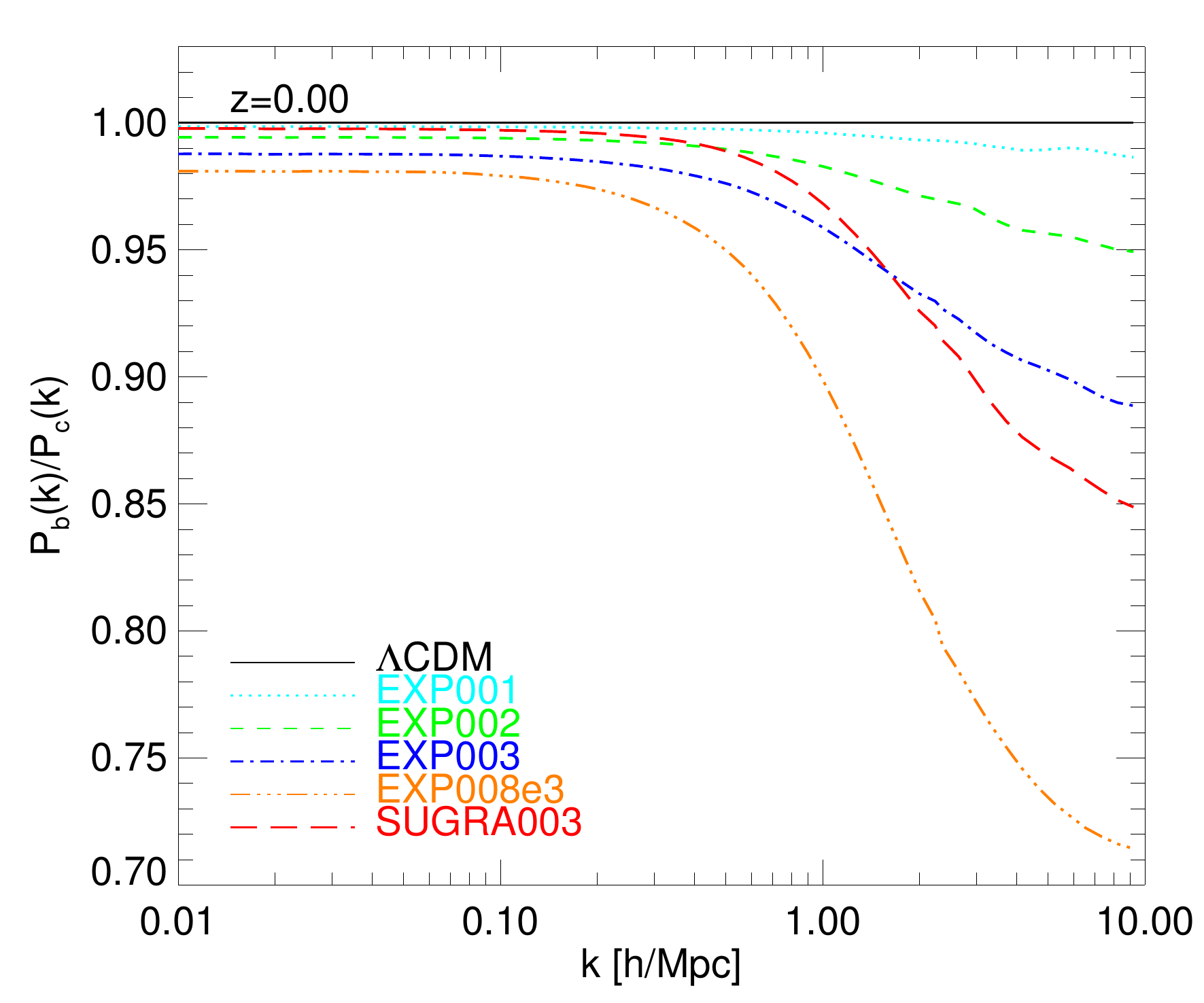}
\caption{The ratio of the baryon to CDM power spectra as a function of scale and at different redshift: $z=1$ (top), $z=0.56$ (middle), $z=0$ (bottom). 
The nonlinear enhancement of the bias clearly develops
as time passes and strongly reduces the baryon fraction of overdense regions at small scales, while at the largest scales the effect never exceeds a few percent.
The growing coupling model EXP008e3 and the bouncing cDE model SUGRA003 show the largest nonlinear bias, which nevertheless is strongly enhanced 
for fundamentally different reasons in the two cases, as explained in the text.}
\label{fig:bias}
\end{figure}
\normalsize

A second characteristic feature of cDE models widely discussed in the literature \citep[see \eg][]{Maccio_etal_2004,Mainini_Bonometto_2006, Baldi_etal_2010} 
arises as a consequence of the different dynamic equations that CDM and baryonic particles 
obey due to their different couplings to the DE scalar field. While baryons are always uncoupled and therefore follow standard Newtonian dynamics,
CDM particles are affected by additional forces (the fifth-force and the extra friction) due to the interaction with the DE scalar field, 
as shown in Eqs.~(\ref{gf_c},\ref{gf_b}). This different dynamical evolution of the two matter
components of the universe determines an offset in the amplitude of density perturbations of the two fluids, that starting from the same amplitude at high redshifts
progressively develop a ``gravitational bias", which has been shown to have a very significant impact on the expected baryon fraction of bound objects as galactic or cluster sized halos \citep[][]{Baldi_etal_2010,Baldi_2011a}.
More specifically, a lower baryon fraction is generically expected in cDE models as compared to $\Lambda $CDM as a consequence of the segregation of baryons 
\citep[first discussed by][]{Mainini_Bonometto_2006}
in the outskirts of collapsed structures determined by the gravitational bias.
The redshift and scale dependence of this effect can be tested by computing the ratio of the nonlinear matter power spectra of the baryonic and CDM components, $P_{b}(k)/P_{c}(k)$.
In Fig.~\ref{fig:bias} we plot this quantity, normalized to the $\Lambda $CDM case, for all the cDE models considered in the present work and at three different redshifts.
As one can see from the figure, the gravitational bias shows a very characteristic scale dependence: while in the linear regime represented by the largest scales a weak and scale-independent offset 
appears for all the models, with a maximum deviation from the standard $\Lambda $CDM value of a few percent, the effect becomes much larger and strongly scale-dependent
when moving to progressively more nonlinear scales, reaching at $z=0$ a suppression of the baryonic power with respect to the CDM power (at $k\sim 10 \, h/$Mpc) of about $1\%$, $5\%$, and $11\%$
for the standard cDE models with constant coupling EXP001 (cyan), EXP002 (green), and EXP003 (blue), respectively.

Particularly interesting are again, however, the cases of the exponential coupling model EXP008e3 (orange) and of the bouncing cDE model SUGRA003 (red). 
These show a much steeper scale dependence in the mildly-nonlinear and nonlinear ranges of scales as compared to the standard cDE models with constant coupling, 
reaching a suppression of the baryonic power with respect to the CDM power
of $\sim 30\%$ and $\sim 15 \%$, respectively, while at large scales the effect has a comparable amplitude to the standard cDE models EXP001-EXP003. 
Interestingly, this enhancement of the gravitational bias at small scales that characterizes the EXP008e3 and the SUGRA003 models has a different origin in the two cases.

The exponential coupling model EXP008e3, in fact, features a sudden enhancement of the growth of CDM perturbations due to the steep increase of the coupling $\beta _{c}(\phi )$ at low redshifts. This 
determines a fast increase of the effective gravitational constant $G_{\rm eff}\equiv G\, \Gamma _{c}(\phi )$ (see Eqs.~\ref{gf_c},\ref{Gamma_c_massless}) 
that applies to CDM particles only, thereby inducing a fast collapse of CDM
overdensities with respect to the corresponding baryonic perturbations. The latter are in fact governed by the standard gravitational constant $G$ and therefore respond with a longer
dynamical timescale to the rapidly growing gravitational potential induced by the CDM collapse. As a consequence, the growth of baryon fluctuations is always delayed with respect to the 
corresponding CDM perturbations, giving rise to the large bias observed at small scales. 
This interpretation is confirmed by following the redshift evolution of the effect: the scale dependence of the bias and the amplitude of the small-scale offset strongly increase
between $z=1$ and $z=0$, in correspondence to the steep increase of the coupling function for this model.
This effect was already investigated by \citet{Baldi_2011a} for a wide range of growing coupling scenarios with a common normalization of linear perturbations at $z=0$. 
We therefore confirm here that the same effect appears also for models with a common normalization of linear perturbations at CMB, although with a significantly larger amplitude
than for the former case. 

The bouncing cDE model SUGRA003, instead, does not show a very strong redshift evolution of the scale dependence of the gravitational bias. In this case, in fact, the origin of the effect is 
substantially different than for the exponential coupling model. 
As discussed in full detail by \citet{Baldi_2011c}, the inversion of the direction of motion of the DE scalar field in the bouncing cDE scenario 
determines a change of sign of the quantity $\dot{\phi }\beta _{c}$, that for the specific model considered here
goes from positive to negative in correspondence to the DE ``bounce" at $z_{\rm inv}\sim 6.8$. 
This implies a change of sign of the extra friction term of Eq.~(\ref{gf_c}) that behaves as a drag term (\ie accelerating
particles along their direction of motion) before the bounce and as a proper friction (\ie decelerating particles) after it.
In the low-redshift universe, then, the SUGRA003 model is characterized by an extra friction term that decelerates CDM particles thereby favoring the collapse of CDM perturbations by removing angular momentum
only from the CDM component. This determines a contraction of CDM perturbations that grow faster than their baryonic counterparts, giving rise to the observed enhancement
of the gravitational bias with respect to standard cDE scenarios.

The level of baryon depletion of density perturbations in the different cDE models will be reflected in the baryonic fraction of massive halos 
\citep[see][]{Baldi_etal_2010,Baldi_2011a} and could provide a further direct observational signature of
cDE scenarios. It is important to notice that the plots are taken from the {\small L-CoDECS}
suite which does not include any hydrodynamical force. The offset between the two components is therefore of pure gravitational origin, due to the effective modification of the gravitational
dynamics of CDM particles in cDE models. Additional physical processes as hydrodynamical pressure, gas cooling, star formation, and feedback mechanisms are therefore expected to modify this picture in a substantial way \citep[see \eg][]{Duffy_etal_2010,vanDaalen_etal_2011,Semboloni_etal_2011}. 
However, it should be stressed here that the gravitational bias described in the present section is a purely dynamical mechanism that starts to develop well before any of such effects can take place, thereby reducing the baryonic reservoir available for astrophysical 
processes. Detailed high-resolution simulations of cDE models including all these additional ingredients will therefore be necessary in order to provide a reliable quantitative determination of the impact of the gravitational bias on observable properties
of bound halos. Nevertheless, the simple estimate discussed in the present Section already provides a clear description of the redshift and scale dependence of the expected bias.

\subsection{The Halo Mass Function}
\label{sec:HMF}

\begin{figure}
\includegraphics[scale=0.45]{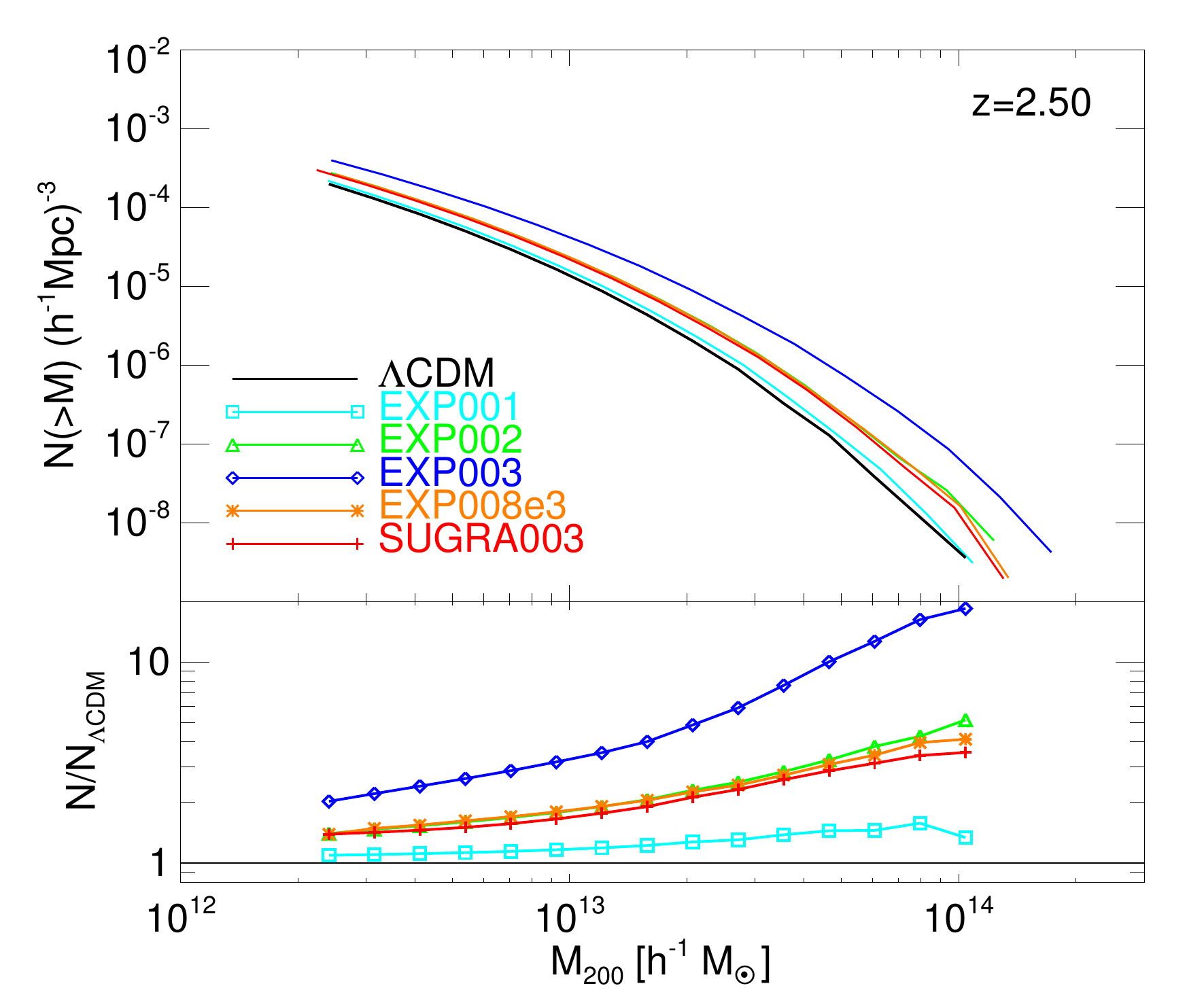}\\
\includegraphics[scale=0.45]{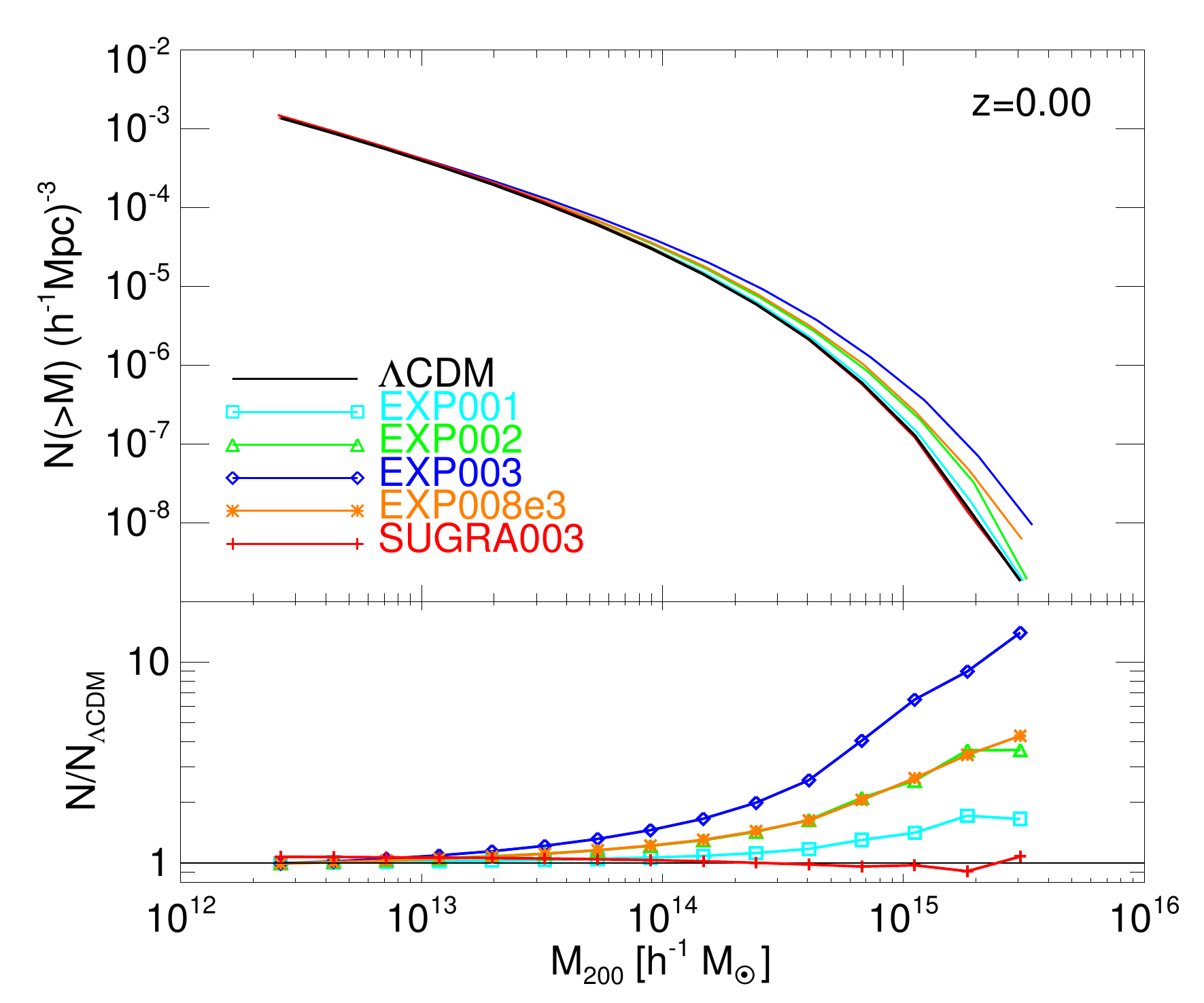}
\caption{The Halo Mass Function of the different cDE models included in the {\small CoDECS} project at $z=2.5$ ({\em upper panel}) and at $z=0$ ({\em lower panel}) as extracted from the halo catalogs of the {\small L-CoDECS} runs based on a Friend-of-Friend identification algorithm. The enhancement of the expected number counts of massive halos is evident in both plots, with a clear mass dependence of the effect. All the models show a significant increase of the number of very massive objects at the present epoch except for the case of the ``Bouncing" cDE model SUGRA003 \citep[see also the more detailed analysisof the Halo Mass Function in cDE models presented in][]{Baldi_2011c,Cui_Baldi_Borgani_2012}}
\label{fig:HMF}
\end{figure}
\normalsize

The enhanced growth of density perturbations and the modified nonlinear dynamics of CDM particles have a significant impact on the expected number of halos as a function of redshift and mass, encoded by the Halo Mass Function (HMF) $N(>M,z)$. The impact of cDE cosmologies on the HMF has been studied in the past by several authors \citep[see \eg][and references therein]{Maccio_etal_2004,Baldi_etal_2010,Baldi_Pettorino_2011,Li_Barrow_2011}, while a specific investigation of the expected number counts of massive clusters at high redshift in the context of the bouncing cDE scenario SUGRA003 discussed in this work has already been carried out by \citet{Baldi_2011c}. Finally, a detailed and comprehensive analysis of the HMF in cDE models based on the {\small CoDECS} simulations has been recently
presented by \citet{Cui_Baldi_Borgani_2012}, resorting both on the Friends-of-Friends (FoF) and on a series of Spherical Overdensity halo samples for all the cosmological models considered in the present work. 

Therefore, we refer the interested reader to the latter work for a thorough discussion on the features of the HMF in cDE scenarios, and we present here only a simple comparison of the expected number counts of massive halos in the different cosmologies considered in the {\small CoDECS} project based on the FoF sample. This has been constructed by running a FoF algorithm with linking length $\ell = 0.2 \times \bar{d}$, where $\bar{d}$ is the mean interparticle separation (see Appendix \ref{appA}) on the CDM particles as primary tracers of the local mass density, and then attaching baryonic particles to the FoF group of their nearest neighbor. In Figure~\ref{fig:HMF} we show the HMF extracted from the {\small L-CoDECS} runs which has been computed using the FoF catalogs at high redshift ($z=2.5$, upper panel) and at the present epoch ($z=0$, lower panel) for all the cosmological models included in the {\small CoDECS} project. The bottom plot in each panel displays the ratio of the number counts with respect to the fiducial $\Lambda $CDM cosmology. 

As one can see from the figures, all the cDE models predict a significant enhancement in the number of halos at high redshift. The effect is visible over the whole mass range covered by our FoF sample, but shows a clear mass dependence with a much stronger enhancement at large masses, reaching a factor of $\sim 20$ at $M\approx 1.0\times 10^{14}$ M$_{\odot }/h$ for the most extreme scenario EXP003. The other models have a weaker -- although still significant -- impact on the expected number counts of massive halos, and it is particularly interesting to notice the enhancement of a factor of $\sim 4$ at $M\approx 1.0\times 10^{14}$ M$_{\odot }/h$ of the bouncing cDE model SUGRA003 \citep[see also][]{Baldi_2011c}.

At low redshift the effect becomes more strongly mass dependent and moves towards larger masses. The most extreme scenario shows an enhancement of a factor of $\sim 15$ at $M\approx 3.0\times 10^{15}$ M$_{\odot }/h$, which reduces to a factor of $\sim 2-5$ for the other models. The only exception is provided by the bouncing cDE model SUGRA003, that shows basically no difference with respect to $\Lambda $CDM at $z=0$ \citep[see again the detailed discussion in][]{Baldi_2011c}.

These results show some of the possible applications of the {\small CoDECS} public halo catalogs that are introduced in the Appendix. The enhancement of the expected number of massive clusters as a function of redshift is one of the distinctive features that characterize cDE scenarios and might provide a direct way to constrain the parameters of the model.

\subsection{The subhalo mass function}
\label{HMF}

\begin{figure}
\includegraphics[scale=0.45]{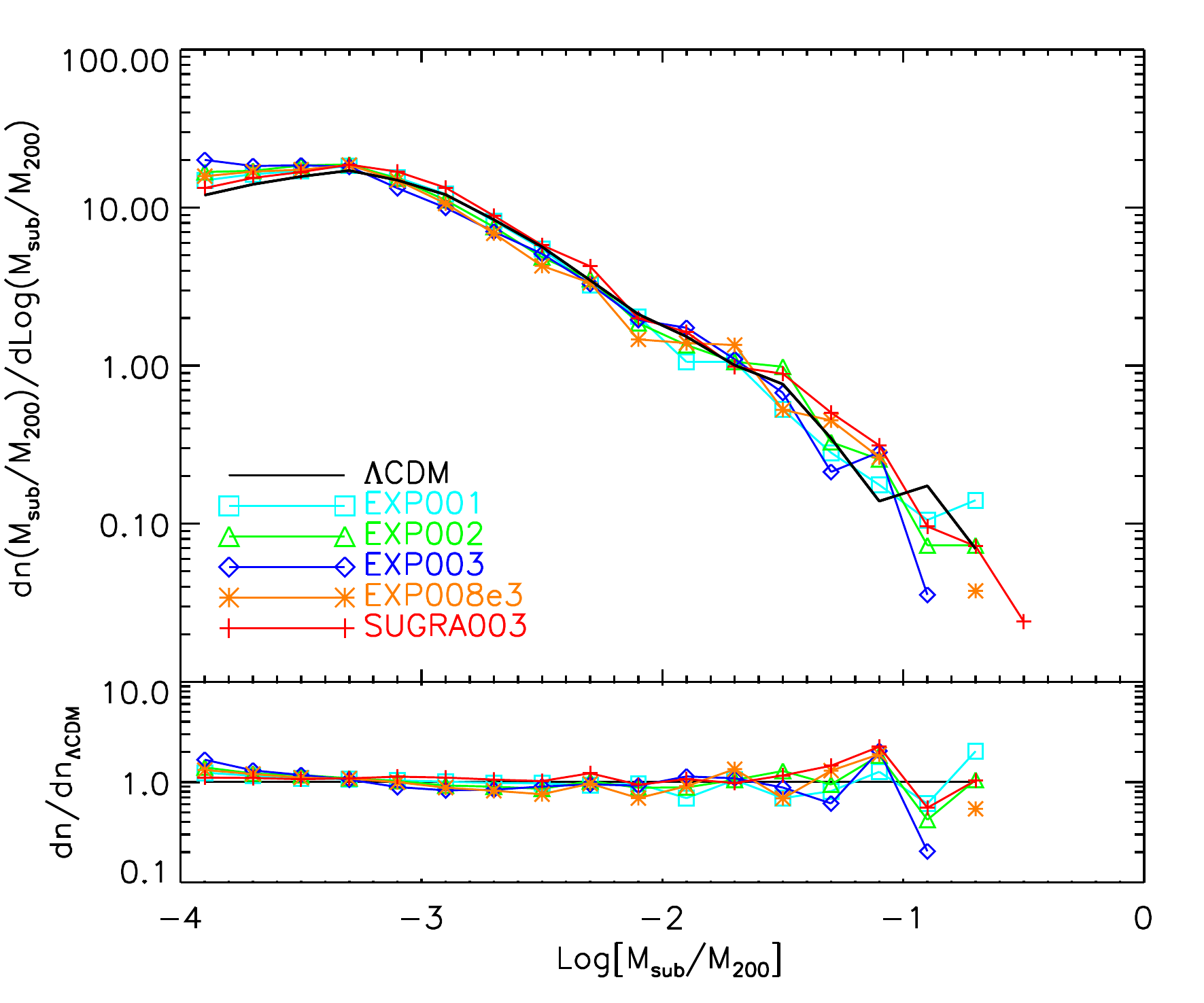}\\
\caption{The Subhalo Mass Function at $z=0$ of all the cDE models included in the {\small CoDECS} project as extracted from the subhalo catalogs of the {\small H-CoDECS} runs based on the {\small SUBFIND} algorithm \citep[][]{Springel2001}. The plot shows the Subhalo Mass Function for host halos with virial mass $M_{200}>5.0\times 10^{12}$ M$_{\odot }/h$. No clear trend appears in the relative abundance of subhalos as a function of the different realizations of DE interactions.}
\label{fig:subHMF}
\end{figure}

As a further example of the possible applications of the {\small CoDECS} public catalogs we present here the subhalo mass function, defined as the number of substructures with a given fractional mass with respect to the virial mass of their host main halo ($M_{\rm sub}/M_{200}$) as a function of the fractional mass itself. We compute this quantity using the subhalo catalogs obtained for all the simulations of the {\small CoDECS} project by means of the {\small SUBFIND} algorithm \citep[][]{Springel2001}, and subsequently binning in logarithmic fractional mass bins the whole sample of substructures belonging to host halos with virial mass above a minimum mass threshold. In Figure~\ref{fig:subHMF} we show the subhalo mass function at $z=0$ as extracted from the subhalo catalogs of the {\small H-CoDECS} simulations. The bottom plot displays the ratio of the subhalo mass function with respect to the fiducial $\Lambda $CDM cosmology. The mass threshold for the main halos has been chosen as $M_{\rm min} = 5\times 10^{12}$ M$_{\odot }/h$.

As one can see from the figure, the mass distribution of halo substructures is not significantly affected by the interaction between DE and CDM, as no clear trend comparable to what observed for the HMF appears in the plot, besides some random scatter at large fractional masses due to the limited number of subhalos in the high-mass bins of the sample. This result shows for the first time how the coupling between DE and CDM does not affect the relative abundance of halos of different masses but rather determines an overall shift in the mass of gravitationally bound structures during the whole assembly process of large halos. A more detailed analysis of the formation process of such halos and of the spatial distribution and dynamical state of subhalos, which goes beyond the scope of the present paper, would be required in order to detect possible footprints of DE interactions in the observable features of halo substructures.

\normalsize

\section{Conclusions}
\label{sec:concl}

We have described the set of cosmological models and the numerical setup adopted for a vast simulations program aimed at investigating
the nonlinear evolution of Coupled Dark Energy cosmologies to an unprecedented level of statistical significance and numerical accuracy.
We refer to such enterprise as the ``{\small CoDECS} project", standing for COupled Dark Energy Cosmological Simulations.
The {\small CoDECS} project includes different types of interacting Dark Energy scenarios and covers a wide range of scales by means 
of two separate suites of cosmological N-body simulations on large scales ({\small L-CoDECS}) and small scales ({\small H-CoDECS}) for each of the models under investigation,
 the latter including also an hydrodynamic treatment for the uncoupled fraction of baryonic particles. 

The range of models considered in the {\small CoDECS} suite includes constant coupling, variable coupling, and bouncing Coupled Dark Energy scenarios, 
and for each of these cosmologies we have described in full detail the background and linear perturbations evolution, derived by numerically integrating the system
of coupled differential dynamic equations from the very early universe to the present time. Linear perturbations have been normalized to the same amplitude at the last scattering surface
in accordance to the latest WMAP7 results. This allows to study, by means of the {\small CoDECS} simulations, 
the evolution of structure formation in the context of different Coupled Dark Energy scenarios that are
all consistent with present CMB constraints. 

We have described in full detail the setup of initial conditions and the numerical features of our different N-body simulations, and provided a visual comparison of
the structures forming in the context of a few selected models as extracted from both the {\small L-CoDECS} and {\small H-CoDECS} runs. Although the
main geometrical structure of the cosmic web appears very similar in all the models, due to the identical phases adopted for the initial conditions generation,
some differences in the overall density contrast and in the distribution and merging history of the smaller structures can be easily identified even by eye.
In particular, for those models that due to the coupling end up with a significantly higher value of $\sigma _{8}$ at $z=0$, the difference in the 
density contrast between highly nonlinear lumps and void regions is particularly evident.\\

We have then moved to a more quantitative analysis of the {\small CoDECS} simulations, by investigating the impact of the different cosmological models on the nonlinear
matter power spectrum. More specifically, we have studied the degeneracy of the effects of the Dark Energy coupling with other cosmological parameters,
in particular with $\sigma _{8}$, by comparing the nonlinear power spectra extracted from each of our simulations to the power spectra computed for a $\Lambda $CDM
cosmology with the same $\sigma _{8}$ of the considered Coupled Dark Energy model by means of the {\small HALOFIT} module implemented in the publicly available code {\small CAMB}.

At linear scales, although a perfect degeneracy was found between Dark Energy coupling and $\sigma _{8}$ at $z=0$, our results have shown for the first time how such degeneracy could
be broken, at least in principle, by the redshift evolution of the linear power spectrum amplitude, that in Coupled Dark Energy decreases more rapidly with redshift as compared to 
a $\Lambda $CDM cosmology with the same $\sigma _{8}$ at $z=0$. 
Even more remarkably, we have shown how the degeneracy can be easily broken, for sufficiently large couplings, even at $z=0$ by moving to nonlinear scales beyond $k\sim 0.3\, h/$Mpc:
the nonlinear effect of the friction term associated to the Dark Energy interaction significantly suppresses the nonlinear power for constant and variable coupling models, while
it strongly increases it for bouncing Coupled Dark Energy scenarios, thereby providing a clear distinction between an interacting Dark Energy and a high-$\sigma _{8}$ $\Lambda $CDM
cosmology.\\

We have also studied the effect of the Dark Energy coupling on the offset between the amplitude of density fluctuations in the CDM and baryonic components, the so called
gravitational bias. Due to the different dynamics determined by the interaction with the Dark Energy scalar field, CDM perturbations are in fact expected to develop faster
with respect to the corresponding baryonic fluctuations, that obey standard Newtonian dynamics. Our analysis has shown, confirming previous results, how the 
offset between the perturbations amplitude in the two components, that is limited to a few percent and does not show any scale dependence at linear scales, gets strongly enhanced 
in the nonlinear regime ($k\gsim 0.3-1.0 \, h/$Mpc), where it acquires also a strong scale dependence.
Such effect has been shown in the past to have a direct impact on the predicted cluster baryon fraction, determining a lower baryonic content of massive CDM halos. 
Here we have highlighted
the cases of growing coupling scenarios and of bouncing Coupled Dark Energy models, for which the effect of baryon segregation at small scales is particularly dramatic,
thereby offering a further signature to observationally constrain such models.\\

Finally, we have provided a few examples of the possible applications of the public {\small CoDECS} halo catalogs by computing the Halo Mass Function from the {\small L-CoDECS} runs at high redshift and at the present epoch and the Subhalo Mass Function from the {\small H-CoDECS} runs at $z=0$. Our results have shown how the relative abundance of massive halos is enhanced in Coupled Dark Energy scenarios with respect to $\Lambda $CDM in a mass- and redshift-dependent way. Furthermore, we have shown for the first time that the mass distribution of substructures within large collapsed objects is not significantly different in Coupled Dark Energy models with respect to the standard $\Lambda $CDM cosmology.
\ \\

The main aim of the present paper is to provide a general description of the cosmological models and of the numerical setup of the {\small CoDECS} project, 
as a a reference for the public release of the simulations outputs. For this reason, we have also described in the Appendix of the paper the main features of
the {\small CoDECS} public database, and the different types of data that are made directly available for public use. More detailed analysis of the {\small CoDECS}
runs will be performed in future works.\\

To conclude, we have performed the largest and most detailed numerical study of Coupled Dark Energy models to date, and provided a general description of our numerical methods and findings.
Our results broadly confirm previous outcomes but significantly extend the statistical significance of the results and the range of models covered by the simulations with respect to previous works, 
also including for
the first time the recently proposed Bouncing Coupled Dark energy scenario. Furthermore, differently from all previous numerical studies of Coupled Dark Energy models,
our {\small CoDECS} simulations are fully consistent with the latest CMB constraints from WMAP7 concerning the normalization of both the background cosmological parameters and 
the amplitude of density perturbations. The simulations outputs, including halo catalogs and nonlinear matter power spectra, are made publicly available through a web database 
and will serve for future analysis of interacting Dark Energy cosmologies.

\section*{Acknowledgments}

This work has been supported by 
the DFG Cluster of Excellence ``Origin and Structure of the Universe''.
All the {\small L-CoDECS} simulations have been performed on the Power6 cluster at the RZG computing centre in Garching and on the SP6 
machine at CINECA (Italy) through an HPC-Europa2 visiting grant. 
The {\small H-CoDECS} simulations have been performed on the ``Sciama" High Performance Compute (HPC) cluster which is supported by 
the ICG, SEPNet and the University of Portsmouth.\\
I am deeply thankful to 
Luca Amendola,
Raul Angulo,
Carlo Baccigalupi,
David Bacon,
Emma Beynon,
Stefano Borgani,
Weiguang Cui,
Gabriella De Lucia,
Veronika Junk,
Kazuya Koyama,
Jounghun Lee,
Federico Marulli,
Lauro Moscardini,
Valeria Pettorino,
and Jochen Weller
for useful discussions at different stages of the development of this work.

\newpage
\appendix
\section{The CoDECS Public Database}
\label{appA}

All the numerical outputs of the {\small CoDECS} project are available for public use. These include both the raw snapshots data of the individual simulations
and some processed quantities as the nonlinear matter power spectra or the halo and sub-halo catalogs. The post-processed data are made directly accessible through a 
web interface at the URL:
\begin{itemize}
\item[] {\bf www.marcobaldi.it/research/CoDECS} 
\end{itemize}
while the unformatted snapshots data, due to their large size, are stored on a separate server
and must be requested directly to the author.
In the present section we will briefly describe which data are available and how post-processing has been performed, while we refer to the specific user's guide
accessible at the {\small CoDECS} website mentioned above for a more specific description of how to interpret the file formats of the processed and raw data.
The numerical features of the two different simulation suites included in the {\small CoDECS} project -- the {\small L-CoDECS} and the {\small H-CoDECS} -- are summarized in Table~\ref{tab:simulation_details}.
\begin{table*}
\begin{tabular}{llcccccc}
\hline
\hline
 & & & & & & & \\
\begin{minipage}{40pt}
\begin{center}
Simulation\\
suite
\end{center}
\end{minipage} & 
\begin{minipage}{30pt}
\begin{center}
Box Size \\ L
\end{center}
\end{minipage}  &  
\begin{minipage}{50pt}
\begin{center}
baryon particle \\ number $N_{b}$
\end{center}
\end{minipage} &
\begin{minipage}{45pt}
\begin{center}
CDM particle \\ number $N_{c}$
\end{center}
\end{minipage} &
\begin{minipage}{50pt}
\begin{center}
baryon particle \\ mass $m_{b}$
\end{center}
\end{minipage} &
\begin{minipage}{45pt}
\begin{center}
CDM particle \\ mass $m_{c}$
\end{center}
\end{minipage} &
\begin{minipage}{45pt}
\begin{center}
Gravitational \\ softening $\epsilon _{g}$
\end{center}
\end{minipage} &
Hydrodynamics \\
 & & & & & & & \\
\hline
 & & & & & & & \\
{\small L-CoDECS} & $1~{\rm Gpc}/h$ & $1024^{3}$ & $1024^{3}$ & $1.17\times 10^{10}$ M$_{\odot}/h$ & $5.84\times 10^{10}$ M$_{\odot}/h$ & $20$ kpc$/h$ & NO \\
 & & & & & & & \\
{\small H-CoDECS} & $80~{\rm Mpc}/h$ & $512^{3}$ & $512^{3}$ & $4.78\times 10^{7}$ M$_{\odot}/h$ & $2.39\times 10^{8}$ M$_{\odot}/h$ & $20$ kpc$/h$ & YES \\
 & & & & & & & \\
\hline
\hline
\end{tabular}
\caption{The numerical parameters of the two suites of runs of the {\small CoDECS} project: the collisionless large-scale suite {\small L-CoDECS} and the hydrodynamic small-scale suite {\small H-CoDECS}.}
\label{tab:simulation_details}
\end{table*}

\subsection{The CoDECS website}
The {\small CoDECS} website provides all the necessary information to access, download, and interpret the {\small CoDECS} data, including the
terms of use and the relevant references. From the URL address given above, one can access a summary page where the main features of all the simulations
of both the {\small L-CoDECS} and {\small H-CoDECS} suite are listed and from where a specific database for each simulation can be accessed. 
Each of the different simulations, in fact, has an individual web page containing all the directly available data for that specific run. All the individual
pages have a standard format that is described in detail below.

\subsection{Background and linear perturbations data}

For each specific model, a number of formatted files containing information about the background and linear perturbations evolution are available.\\
\ \\
The background files:
\begin{itemize}
\item[] {\em model}\textunderscore CoDECS-background.dat
\end{itemize}
provide the redshift evolution of the relative energy fraction of all the different components of the universe $\Omega _{c}$, $\Omega _{b}$,
$\Omega _{r}$, $\Omega _{\rm kin}$, $\Omega _{\rm pot}$, where the latter two quantities represent the fractional energy density of the kinetic and potential
contributions to the DE density $\Omega _{\phi }$, respectively. In addition, the Hubble function $H$ (normalized to $0.1$ at the present time), the coupling 
function $\beta _{c}$, the CDM
particle mass $m_{c}$ (normalized to its present-day value), and the DE equation of state parameter $w_{\phi }$ are included in the file. All the plots in
Fig.~\ref{fig:background} have been obtained using information which is included in these files.\\
\ \\
Other three files:
\begin{itemize}
\item[] {\em model}\textunderscore CoDECS-growthfactor\textunderscore CDM.dat
\item[] {\em model}\textunderscore CoDECS-growthfactor\textunderscore Baryons.dat
\item[] {\em model}\textunderscore CoDECS-growthfactor\textunderscore all.dat
\end{itemize}
provide information on the evolution of the linear growth factor for CDM, baryons, and the total matter component, respectively, including the evolution
of the quantity $D_{+}/a$ for these three different cases. Both plots in Fig.~\ref{fig:linear} were obtained using data included in these files.\\
\ \\
Finally, the linear matter power spectrum used to set up the initial conditions of the simulations and normalized to the amplitude at $z=0$ defined by $\sigma _{8}$ for each specific model is also provided, both
in terms of the dimensionless power spectrum $\Delta ^{2}(k)$ and of the dimensional power spectrum $P(k)\equiv 2\pi ^{2}\Delta ^{2}(k)/k^{3}$, in the files:
\begin{itemize}
\item[] {\em model}\textunderscore CoDECS-linear\textunderscore power.dat
\end{itemize}

\subsection{Non-linear matter power spectra}

A series of links provides access to formatted files named:
\begin{itemize}
\item[] {\em model}\textunderscore CoDECS-power\textunderscore CDM\textunderscore XXX.dat
\item[] {\em model}\textunderscore CoDECS-power\textunderscore Baryons\textunderscore XXX.dat
\item[] {\em model}\textunderscore CoDECS-power\textunderscore all\textunderscore XXX.dat
\end{itemize}
for the nonlinear power spectra of CDM, baryons, and total matter, respectively, as extracted from the simulations at different redshifts. The number `XXX' indicates the 
snapshot number, and a direct correspondence between snapshot number and redshift is provided for each link.
As mentioned above, the power spectra have been computed at scales larger than the Nyquist frequency of each simulation by performing a CIC (Cloud-In-Cell)
assignment to a grid of the same size of the PM grid, \ie $1024^3$ and $512^3$ for the {\small L-CoDECS} and {\small H-CoDECS} sets of simulations, respectively.
For Fourier modes beyond the Nyquist frequency the power spectrum is computed with the folding method \citep[][]{Jenkins_etal_1998,Colombi_etal_2009} and then
smoothly interpolated to the large-scale power in the region of overlap, in order to provide a single estimate of the nonlinear power up to modes where the shot-noise
is found to be $10\%$ of the measured power, at which we cut our power spectrum estimate. Also in this case, the files include both the dimensionless and dimensional
power spectra, $\Delta ^{2}(k)$ and $P(k)$, respectively.

\subsection{Groups and SubGroups catalogs}

For all the {\small CoDECS} simulations, halos have been identified by means of a Friends-of-Friends (FoF) algorithm with linking length $\ell = 0.2 \times \bar{d}$,
where $\bar{d}$ is the mean interparticle separation. This procedure has been applied to the particle distribution in the simulations for a large number of redshifts
by linking the CDM particles as primary tracers of the local mass density, and then attaching baryonic particles to the FoF group of their nearest neighbor.
For each FoF halo we have computed a series of global properties as the total number of member particles, the total halo mass, the center-of-mass position and velocity.
All this information, together with the number of gravitationally bound substructures identified
within each single FoF halo, is included in the Groups files of the {\small CoDECS} database, that are grouped in archive files named:
\begin{itemize}
\item[] {\em model}\textunderscore CoDECS-Groups\textunderscore XXX.tar .
\end{itemize}
Each archive contains 512 files for the {\small L-CoDECS} simulations and 128 files for the {\small H-CoDECS} simulations, corresponding to the individual
files written by each processor in the different runs.\\

The identification of the halo substructures is made by means of the {\small SUBFIND} algorithm \citep[][]{Springel2001} and a large number of properties of the individual 
substructures identified in each FoF halo are computed. These include the location of the local gravitational potential minimum, the number of particles belonging
to the substructure, the total mass of the substructure, as well as its peculiar velocity and its specific angular momentum. 
The latter is defined as:
\begin{equation}
\vec{s} \equiv \frac{\sum_{i}m_{i} \left( \vec{r}_{i} \times \vec{v}_{i}\right) }{\sum_{i}m_{i}}
\end{equation}
where $i$ runs over the particles bound to the subhalo, $\vec{r}_{i}$ is the position of the $i$-th particle with respect to the potential minimum of the
subhalo in physical coordinates, and $\vec{v}_{i}$ is its physical peculiar velocity.
For the main substructure in each FoF halo, we also compute the virial radius $R_{200}$, defined as the radius of a sphere centered in the minimum of the gravitational 
potential of the substructure that encloses a mean density $200$ times larger than the cosmic critical density, and the corresponding enclosed virial mass $M_{200}$.
These data are included in the SubGroups files directly available from the {\small CoDECS} web database, which are also grouped in 
archive files named:
\begin{itemize}
\item[] {\em model}\textunderscore CoDECS-SubGroups\textunderscore XXX.tar .
\end{itemize}
Detailed information on the format and the units
of the Groups and SubGroup data is provided in the {\small CoDECS} user's guide directly downloadable from the {\small CoDECS} web site.

\bibliographystyle{mnras}
\bibliography{baldi_bibliography}

\label{lastpage}

\end{document}